\documentclass[sigconf, nonacm]{acmart}

\AtBeginDocument{
  }

\usepackage[linesnumbered,ruled,vlined]{algorithm2e}
\usepackage{xspace}
\usepackage{graphicx}
\usepackage{enumitem}
\def\oursolution{SO-COD\xspace}
\def\newsolution{ML-SO-COD\xspace}
\def\lsn{\frac{6}{\epsilon}}
\def\xinorm{\left\|\boldsymbol{x}_i\right\|_2}
\def\yinorm{\left\|\boldsymbol{y}_i\right\|_2}

\newcommand{\htitle}[1]{\vspace{1mm} \noindent \textbf{#1}}

\definecolor{LightSteelBlue}{RGB}{213,229,255}
\newtheorem{definition} {Definition}
\newtheorem{example} {Example}
\newtheorem{theorem} {Theorem}
\newtheorem{lemma} {Lemma}

\SetKwInput{KwInput}{Input}
\SetKwInput{KwOutput}{Output}

\begin{document}

%\fancyhead{}

\title{Optimal Approximate Matrix Multiplication over Sliding Window}

\author{Haoming Xian}
\affiliation{
  \institution{The Chinese University of Hong Kong}
  \city{Hong Kong SAR}
    \country{China}
}
\email{hmxian@se.cuhk.edu.hk}

\author{Qintian Guo}
\affiliation{
  \institution{The Hong Kong University of Science and Technology}
  \city{Hong Kong SAR}
    \country{China}
}
\email{qtguo@ust.hk}

\author{Jun Zhang}
\affiliation{
  \institution{Bitlink Capital Limited}
  \city{Hong Kong SAR}
    \country{China}
}
\email{zj@bitlink.capital}

\author{Sibo Wang}
\affiliation{
  \institution{The Chinese University of Hong Kong}
  \city{Hong Kong SAR} 
    \country{China}
}
\email{swang@se.cuhk.edu.hk}

\begin{abstract}
\label{sec:sec-abstract}

Matrix multiplication is a core operation in numerous applications, yet its exact computation becomes prohibitively expensive as data scales, especially in streaming environments where timeliness is critical. In many real-world scenarios, data arrives continuously, making it essential to focus on recent information via sliding windows. While existing approaches offer approximate solutions, they often suffer from suboptimal space complexities when extended to the sliding-window setting.

In this work, we introduce SO-COD, a novel algorithm for approximate matrix multiplication (AMM) in the sliding-window streaming setting, where only the most recent data is retained for computation. Inspired by frequency estimation over sliding windows, our method tracks significant contributions—referred to as ``snapshots''—from incoming data and efficiently updates them as the window advances.  Given matrices \(\boldsymbol{X} \in \mathbb{R}^{d_x \times n}\) and \(\boldsymbol{Y} \in \mathbb{R}^{d_y \times n}\) for computing \(\boldsymbol{X} \boldsymbol{Y}^T\), we analyze two data settings. In the \emph{normalized} setting, where each column of the input matrices has a unit \(L_2\) norm, SO-COD achieves an optimal space complexity of \( O\left(\frac{d_x+d_y}{\epsilon}\right) \). In the \emph{unnormalized} setting, where the square of column norms vary within a bounded range \([1, R]\), we show that the space requirement is \( O\left(\frac{d_x+d_y}{\epsilon}\log R\right) \), which matches the theoretical lower bound for an \(\epsilon\)-approximation guarantee. Extensive experiments on synthetic and real-world datasets demonstrate that SO-COD effectively balances space cost and approximation error, making it a promising solution for large-scale, dynamic streaming matrix multiplication.

\end{abstract}

\maketitle

\section{Introduction}
\label{sec:sec-Introduction}

Matrix multiplication is a core operation across machine learning, data analysis, signal processing, and computer graphics. However, as data volumes grow, exact matrix multiplication becomes increasingly expensive, especially in large-scale scenarios demanding real-time or near-real-time processing. To address these challenges, approximate matrix multiplication (AMM) has emerged as an attractive alternative, offering substantial reductions in computational and storage overhead while preserving high accuracy.

In real-world applications, data often arrives in a streaming fashion, as exemplified by social media analytics (where posts or tweets are continuously generated), user behavior analysis (where search queries or web content are continuously updated), and financial market monitoring (where new transactions or price quotes are streamed in). This continuous flow of data has spurred interest in extending Approximate Matrix Multiplication (AMM) algorithms to streaming models \cite{MrouehMG17, YeLZ16}. Among various techniques, from sampling \cite{DrineasKM06} and random projections \cite{CohenNW16,MagenZ11,Sarlos06} to hashing \cite{ClarksonW13}, Co-occurring Directions (COD) \cite{MrouehMG17} has gained prominence for its robust approximation guarantee and strong empirical performance. 

However, a key limitation of traditional streaming methods is that they treat all incoming data uniformly, without prioritizing more recent information. In many real-world applications, the most valuable insights stem from the most recent data, necessitating a focus on a sliding window of the most current columns rather than retaining the entire historical dataset. For instance, in social media analysis, matrix \( \boldsymbol{X} \) could represent a stream of recent tweets, while matrix \(\boldsymbol{Y} \) could represent user interests. The multiplication \( \boldsymbol{X}  \boldsymbol{Y}^T \) captures the relevance of each tweet to user preferences. Similarly, in user behavior analysis, matrix \( \boldsymbol{X} \) might represent recent search queries, and matrix \( \boldsymbol{Y} \) could represent the relevance of advertisements or content. By using a sliding window, the approach prioritizes recent data, ensuring that the most current user activities influence the recommendations. Thus, matrix multiplication over a sliding window in a streaming setting is a natural choice, as it aligns with the data evolving nature and emphasizes the timeliness of insights, which is crucial for real-world applications.

To address this challenge, a line of research has focused on developing more efficient sketches that use less space while maintaining the same approximation guarantees. For example, by integrating techniques for sliding-window sampling with sampling-based matrix multiplication (see \cite{efraimidis2006weighted, drineas2006fast, babcock2001sampling}), we can discard outdated samples and draw new ones from the most recent rows or columns of the input matrix. However, the space cost of the sampling-based method remains relatively high, scaling with \( O\left(\frac{1}{\epsilon^2} \log N\right) \), where \( N \) is the number of rows and columns involved in the multiplication, and \( \epsilon \) is the relative error guarantee for approximate matrix multiplication (refer to Def.\ \ref{def:amm}). 
Yao et al.~\cite{YaoLCWC24} present two algorithms, EH-COD and DI-COD, which reduce the dependence on \( N \). However, they still incur a space cost that involves either a quadratic dependence on \( \frac{1}{\epsilon} \) or an extra factor of \( \log^2\left(\frac{1}{\epsilon}\right) \).
Thus, although progress has been made, the space complexity of existing sketching methods remains suboptimal, indicating room for further improvement.

Motivated by these limitations, we introduce {\em \oursolution} (\underline{S}pace \underline{O}ptimal \underline{C}o-\underline{O}ccuring \underline{D}irections), a novel algorithm specifically designed for approximate matrix multiplication in the sliding window setting. By maintaining a compact yet representative summary of the most recent data, {\oursolution} not only achieves optimal space complexity—matching the best-known theoretical bounds from the streaming literature—but also effectively meets the demands of time-sensitive applications.

At a high level, our approach draws inspiration from the classic sliding window frequency estimation problem \cite{LeeT06}, where one estimates the frequency of individual elements over the most recent \(N\) items by recording snapshot events when an element’s count reaches a threshold and expiring outdated snapshots as the window slides. In our work, we extend this snapshot concept to the significantly more challenging task of approximate matrix multiplication (AMM) for the matrices \(\boldsymbol{X} \in \mathbb{R}^{d_x \times n}\) and \(\boldsymbol{Y} \in \mathbb{R}^{d_y \times n}\). Rather than tracking individual elements, we register the key directions that contribute substantially to the product \(\boldsymbol{X}\boldsymbol{Y}^T\) as snapshots, and we expire these snapshots once they fall outside the sliding window. This extension is non-trivial since matrix multiplication involves complex interactions that far exceed the simplicity of frequency estimation. It is crucial to emphasize that, although Yin et al.\ \cite{YinWLWZHL24} also utilize a \(\lambda\)-snapshot framework, their work is confined to approximating covariance matrices (i.e., \(\boldsymbol{A}\boldsymbol{A}^T\)), which represents only a special case of the AMM problem. In contrast, our method is designed to address the full generality of AMM, overcoming challenges that their approach cannot.

We consider two settings. First, when every column of \(\boldsymbol{X}\) and \(\boldsymbol{Y}\) is normalized (i.e., each column has an \(L_2\) norm of 1), our method requires space \(O\left(\frac{d_x+d_y}{\epsilon}\right)\). We then extend our solution to the unnormalized case, where the squared \(L_2\) norm of each column lies within \([1, R]\), and demonstrate that the space cost increases to \(O\left(\frac{d_x+d_y}{\epsilon}\log R\right)\). In both cases, we prove that these bounds are tight by matching known space lower bounds for an \(\epsilon\)-approximation guarantee (see Def.\ \ref{def:amm}). Table \ref{tab:table-comparsion} summarizes the space complexity of {\oursolution} compared to existing solutions, clearly demonstrating its advantage. We evaluate the performance of {\oursolution} with extensive experiments on both synthetic and real-world datasets, showing its accuracy and space efficiency in comparison to existing methods. The results confirm that {\oursolution} effectively balances the space cost and approximation errors, making it a practical choice for large-scale streaming matrix multiplication in dynamic environments.

\section{Preliminaries}
\label{sec:sec-Preliminary}

\begin{table}[t]
  \caption{
  Space complexity of various approximate matrix multiplication algorithms in the sliding-window setting. 
  %Note that the final complexity is scaled by \(O(d_x + d_y)\).
  %Space complexity for various AMM algorithms under the sliding-window setting.
  }
  \vspace{-2mm}
  \label{tab:table-comparsion}
  \renewcommand{\arraystretch}{1.5}
  \begin{tabular}{|c|c|c|}
    \hline
    Algorithm & Normalized model & Unnormalized model \\
    \hline
    Sampling \cite{efraimidis2006weighted, drineas2006fast,babcock2001sampling} & $O(\frac{d_x + d_y}{\epsilon^2}\log{N})$ & $O\left(\frac{d_x + d_y}{\epsilon^2}\log{NR}\right)$ \\
    \hline
    DI-COD \cite{YaoLCWC24} & $O(\frac{d_x + d_y}{\epsilon}\log^2{\frac{1}{\epsilon}})$ & $O\left(\frac{(d_x + d_y)\cdot R}{\epsilon}\log^2{\frac{R}{\epsilon}}\right)$\\
    \hline
    EH-COD \cite{YaoLCWC24} & $O(\frac{d_x + d_y}{\epsilon^2}\log{\epsilon N})$ & $O\left(\frac{d_x + d_y}{\epsilon^2}\log{\epsilon NR}\right)$ \\
    \hline
    % DI-SCOD \cite{YaoLCWC24} & $O(\frac{d_x + d_y}{\epsilon}\log^2{\frac{1}{\epsilon}})$ & $O\left(\frac{(d_x + d_y)\cdot R}{\epsilon}\log^2{\frac{R}{\epsilon}}\right)$\\
    % \hline
    % EH-SCOD \cite{YaoLCWC24} & $O(\frac{d_x + d_y}{\epsilon^2})$ & $O\left(\frac{d_x + d_y}{\epsilon^2}\log{R}\right)$ \\
    % \hline
    {\oursolution} (Ours) & $O(\frac{d_x + d_y}{\epsilon})$ & $O\left(\frac{d_x + d_y}{\epsilon}\log{R}\right)$ \\
    \hline
\end{tabular}
\vspace{-1mm}
\end{table}

\subsection{Notations and Basic Concepts}

In this paper, we use bold uppercase letters to denote matrices (e.g., $\boldsymbol{A}$), bold lowercase letters to denote vectors (e.g., $\boldsymbol{x}$), and lowercase letters to denote scalars (e.g., $w$).

Given a real matrix \(\boldsymbol{A} \in \mathbb{R}^{n\times m}\), its condensed singular value decomposition (SVD) is given by \(\boldsymbol{A} = \boldsymbol{U} \boldsymbol{\Sigma} \boldsymbol{V}^\top\), where \(\boldsymbol{U} \in \mathbb{R}^{n \times r}\) and \(\boldsymbol{V} \in \mathbb{R}^{m \times r}\) are matrices with orthonormal columns, and \(\boldsymbol{\Sigma} = \operatorname{diag}(\sigma_1, \sigma_2, \dots, \sigma_r)\) is a diagonal matrix containing the singular values of \(\boldsymbol{A}\) arranged in non-increasing order, i.e., \(\sigma_1 \geq \sigma_2 \geq \cdots \geq \sigma_r > 0\), with \(r\) denoting the rank of \(\boldsymbol{A}\). We define several matrix norms as follows: the Frobenius norm of \(\boldsymbol{A}\) is given by 
\(
\|\boldsymbol{A}\|_F = \sqrt{\sum_{i,j} A_{i,j}^2} = \sqrt{\sum_{i=1}^r \sigma_i^2},
\)
the spectral norm is 
\(
\|\boldsymbol{A}\|_2 = \sigma_1,
\)
and the nuclear norm is 
\(
\|\boldsymbol{A}\|_* = \sum_{i=1}^r \sigma_i.
\)
For a matrix \(\boldsymbol{A}\), we denote its \(i\)-th column by \(\boldsymbol{a}_i\); hence, if \(\boldsymbol{A} \in \mathbb{R}^{n \times l}\) and \(\boldsymbol{B} \in \mathbb{R}^{m \times l}\), then their product can be expressed as: 
\(
\boldsymbol{A}\boldsymbol{B}^T = \sum_{i=1}^l \boldsymbol{a}_i \boldsymbol{b}_i^T.
\)
Finally, \(\boldsymbol{I}_n\) denotes the \(n \times n\) identity matrix and \(\boldsymbol{0}^{n \times m}\) denotes the \(n \times m\) zero matrix.

\subsection{Problem Definition}
The Approximate Matrix Multiplication (AMM) problem has been extensively studied in the literature \cite{YeLZ16,MrouehMG17,drineas2006fast,FrancisR22}. Given two matrices \(\boldsymbol{X} \in \mathbb{R}^{d_x \times n}\) and \(\boldsymbol{Y} \in \mathbb{R}^{d_y \times n}\), the AMM problem aims to obtain two smaller matrices \(\boldsymbol{B}_X \in \mathbb{R}^{d_x \times l}\) and \(\boldsymbol{B}_Y \in \mathbb{R}^{d_y \times l}\), with \(l \ll n\), so that
\(
\left\|\boldsymbol{X}\boldsymbol{Y}^\top - \boldsymbol{B}_X \boldsymbol{B}_Y^\top\right\|_2
\) is sufficiently small. Next, we formally define the problem of AMM over a Sliding Window.

\vspace{-1mm}
\begin{definition}[AMM over a Sliding Window]\label{def:amm}
Let \(\{(\boldsymbol{x}_t, \boldsymbol{y}_t)\}_{t \ge 1}\) be a sequence of data items (a data stream), where for each time \(t\) we have \(\boldsymbol{x}_t \in \mathbb{R}^{d_x}\) and \(\boldsymbol{y}_t \in \mathbb{R}^{d_y}\). For a fixed window size \(N\) and for any time \(T \ge N\), define the sliding window matrices
\[
\boldsymbol{X}_W = \begin{bmatrix} \boldsymbol{x}_{T-N+1} & \boldsymbol{x}_{T-N+2} & \cdots & \boldsymbol{x}_T \end{bmatrix} \in \mathbb{R}^{d_x \times N},\]
\[
\boldsymbol{Y}_W = \begin{bmatrix} \boldsymbol{y}_{T-N+1} & \boldsymbol{y}_{T-N+2} & \cdots & \boldsymbol{y}_T \end{bmatrix} \in \mathbb{R}^{d_y \times N}.
\]
A streaming algorithm (or matrix sketch) \(\kappa\) is said to yield an \(\epsilon\)-approximation for AMM over the sliding window if, at every time \(T \ge N\), it outputs matrices \(\boldsymbol{A}_W \in \mathbb{R}^{d_x \times m}\) and \(\boldsymbol{B}_W \in \mathbb{R}^{d_y \times m}\) (with \(m \le N\), typically \(m \ll N\)) satisfying
\[
\left\|\boldsymbol{X}_W \boldsymbol{Y}_W^\top - \boldsymbol{A}_W \boldsymbol{B}_W^\top\right\|_2 \le \epsilon\, \|\boldsymbol{X}_W\|_F \, \|\boldsymbol{Y}_W\|_F.
\]
That said, the product \(\boldsymbol{A}_W \boldsymbol{B}_W^\top\) produced by the sketch \(\kappa\) approximates the true product \(\boldsymbol{X}_W \boldsymbol{Y}_W^\top\) with a spectral norm error that is at most an \(\epsilon\)-fraction of the product of the Frobenius norms of \(\boldsymbol{X}_W\) and \(\boldsymbol{Y}_W\).
\end{definition}
\vspace{-1mm}

In \cite{WeiLLSDW16}, it is shown that one may assume without loss of generality that the squared norms of the columns of \(\boldsymbol{X}\) and \(\boldsymbol{Y}\) are normalized to lie in the intervals \([1, R_x]\) and \([1, R_y]\), respectively, which is a mild assumption. We denote \(R = \max(R_x, R_y)\).

Next, we present two key techniques that underpin our solution: the Co-occurring Directions (COD) algorithm \cite{MrouehMG17} and the \(\lambda\)-snapshot method \cite{LeeT06}. While each technique was originally developed for a different problem, we show later that their careful and nontrivial integration yields a solution with optimal space to gain $\epsilon$-approximation guarantee.

\subsection{Co-Occurring Directions}

Co-occurring Directions (COD) \cite{MrouehMG17} is a deterministic streaming algorithm for approximate matrix multiplication. Given matrices \(\boldsymbol{X} \in \mathbb{R}^{d_x \times n}\) and \(\boldsymbol{Y} \in \mathbb{R}^{d_y \times n}\) whose columns arrive sequentially, COD maintains sketch matrices \(\boldsymbol{B}_X \in \mathbb{R}^{d_x \times l}\) and \(\boldsymbol{B}_Y \in \mathbb{R}^{d_y \times l}\) (with \(l \ll n\)) so that 
\(
\boldsymbol{X}\boldsymbol{Y}^\top \approx \boldsymbol{B}_X \boldsymbol{B}_Y^\top.
\)
Algorithm \ref{alg:cod} shows the pseudo-code of COD algorithm. In essence, each incoming column pair \((\boldsymbol{x}_i, \boldsymbol{y}_i)\) is inserted into an available slot in the corresponding sketch (Lines 3-4). When a sketch becomes full, a compression step is performed (Lines 5-12): the sketches are first orthogonalized via QR decomposition (Lines 6-7); then, an SVD is applied to the product of the resulting factors (Line 8); finally, the singular values are shrunk by a threshold \(\delta\) (typically chosen as the \(\ell/2\)-th singular value) to update the sketches (Lines 9-12). This process effectively discards less significant directions while preserving the dominant correlations, thereby controlling the approximation error. 

\begin{algorithm}[t]
    \caption{Co-occurring Directions (COD)}
    \label{alg:cod}
    \DontPrintSemicolon
    \KwInput{\(\boldsymbol{X}\in \mathbb{R}^{d_x\times n}\), \(\boldsymbol{Y}\in \mathbb{R}^{d_y\times n}\), sketch size \(l\)}
    \KwOutput{\(\boldsymbol{A}\in \mathbb{R}^{d_x\times l}\) and \(\boldsymbol{B}\in \mathbb{R}^{d_y\times l}\)}
    
    Initialize \(\boldsymbol{A} \leftarrow \boldsymbol{0}^{d_x\times l}\) and \(\boldsymbol{B} \leftarrow \boldsymbol{0}^{d_y\times l}\)\;
    
    \For{\(i = 1, \dots, n\)}{
        Insert \(\boldsymbol{x}_i\) into an empty column of \(\boldsymbol{A}\)\;
        
        Insert \(\boldsymbol{y}_i\) into an empty column of \(\boldsymbol{B}\)\;
        
        \If{\(\boldsymbol{A}\) or \(\boldsymbol{B}\) is full}{
            \((\boldsymbol{Q}_x, \boldsymbol{R}_x) \leftarrow \text{QR}(\boldsymbol{A})\)\;
            
            \((\boldsymbol{Q}_y, \boldsymbol{R}_y) \leftarrow \text{QR}(\boldsymbol{B})\)\;
            
            \([\boldsymbol{U},\boldsymbol{\Sigma},\boldsymbol{V}] \leftarrow \text{SVD}(\boldsymbol{R}_x \boldsymbol{R}_y^\top)\)\;
            
            \(\delta \leftarrow \sigma_{l/2}(\boldsymbol{\Sigma})\), \(\hat{\boldsymbol{\Sigma}} \leftarrow \max(\boldsymbol{\Sigma} - \delta\,\boldsymbol{I}_l, \boldsymbol{0})\)\;
            
            Update \(\boldsymbol{A} \leftarrow \boldsymbol{Q}_x\,\boldsymbol{U}\,\sqrt{\hat{\boldsymbol{\Sigma}}}\)\;
            Update \(\boldsymbol{B} \leftarrow \boldsymbol{Q}_y\,\boldsymbol{V}\,\sqrt{\hat{\boldsymbol{\Sigma}}}\)\;
        }
    }
    
    \Return \(\boldsymbol{A}, \boldsymbol{B}\)
\end{algorithm}

\vspace{-1mm}
\begin{theorem} \label{thm:cod}
The output of Co-occurring Directions (Algorithm~\ref{alg:cod}) provides correlation sketch matrices \((\boldsymbol{B}_X \in \mathbb{R}^{d_x \times l}, \boldsymbol{B}_Y \in \mathbb{R}^{d_y \times l})\) for \((\boldsymbol{X} \in \mathbb{R}^{d_x \times n}, \boldsymbol{Y} \in \mathbb{R}^{d_y \times n})\), where \(l \leq \min(d_x, d_y)\), satisfying:
\[
\left\|\boldsymbol{X} \boldsymbol{Y}^\top - \boldsymbol{B}_X \boldsymbol{B}_Y^\top\right\|_2 \leq \frac{2\|\boldsymbol{X}\|_F \|\boldsymbol{Y}\|_F}{l}.
\]
Algorithm~\ref{alg:cod} runs in \(O(n(d_x + d_y)l)\) time using \(O((d_x + d_y)l)\) space.
\end{theorem}

\subsection{$\boldsymbol{\lambda}$-Snapshot Method}

We explain the key idea of the $\lambda$-snapshot method \cite{LeeT06} by beginning with a bit stream
\(
f = \{b_1, b_2, b_3, \dots\}
\)
where 
\(
b_i \in \{0,1\},
\)
and the goal is to approximate the number of 1-bits in a sliding window. The \(\lambda\)-snapshot method achieves this by “sampling” every \(\lambda\)-th 1-bit. That is, if we index the 1-bits in order of appearance, the \(\lambda\)-th, \(2\lambda\)-th, \(3\lambda\)-th, etc., are stored. The stream is conceptually divided into blocks of \(\lambda\) consecutive positions (called \(\lambda\)-blocks), and a block is deemed \emph{significant} if it intersects the current sliding window and contains at least one sampled 1-bit. The algorithm maintains a \(\lambda\)-counter that tracks: {\em (i)} a queue \(Q\) holding the indices of significant \(\lambda\)-blocks; {\em (ii)} a counter \(\ell\) recording the number of 1-bits seen since the last sample; and {\em (iii)} auxiliary variables for the current block index and the offset within that block. When a new bit arrives, the offset is incremented and any block that no longer falls within the sliding window is removed from \(Q\). If the offset reaches \(\lambda\), the block index is incremented. Upon encountering a 1-bit, \(\ell\) is incremented; when \(\ell\) reaches \(\lambda\), that 1-bit is sampled (i.e., \(\ell\) is reset to 0) and the current block index is \underline{\em registered} and appended to \(Q\). The estimated count of 1-bits in the current window \(W_p\) is given by
\(
v(S) = |Q|\lambda + \ell.
\)
If the true count is \(m\), then it is guaranteed that:
\(
m \le v(S) \le m + 2\lambda,
\)
so that the error is bounded by \(2\lambda\). As the window slides, blocks falling entirely out of range are removed from \(Q\), ensuring that the estimate \(v(S) = |Q|\lambda + \ell\) remains valid with an error of at most \(2\lambda\). 

An example of how the $\lambda$-snapshot method works to count the number of 1-bits in a sliding window is provided in Appendix \ref{app:examples}.

To support frequency estimation over sliding window, a naive approach is to apply the $\lambda$-snapshot method to every distinct element in the stream. For any given element $e$, the stream can be represented as a bit stream, where each new bit is set to 1 if the stream element equals $e$ and 0 otherwise. However, maintaining a separate $\lambda$-snapshot structure for each element would lead to unbounded space usage. Lee et al.\ \cite{LeeT06} show that by maintaining only $O(1/\epsilon)$ such $\lambda$-snapshot structures and combine with the well known frequency estimation algorithm MG-sketch \cite{MisraG82}, an $\epsilon$-approximation for the frequency of each element can be achieved while using just $O(1/\epsilon)$ space. Although there are $O(1/\epsilon)$ $\lambda$-snapshot structures, they collectively track a stream containing at most $N$ ones, ensuring that the overall space cost remains bounded. For each element $e$, let $f(e)$ be its true frequency and $\hat{f}(e)$ be the estimated frequency derived using the $\lambda$-snapshot method, then
\(
f(e)-\hat{f}(e) \leq \epsilon N.
\)

\section{Our Solution}
\label{sec:sec-prim}

\begin{algorithm}[t]
    \caption{\oursolution: Initialize(\(d_x,d_y,l,N,\theta\))}
    \label{alg:initialize}
    \DontPrintSemicolon
    \KwInput{
      \(d_x,d_y\): dimensions of the input spaces for \(\boldsymbol{X}\) and \(\boldsymbol{Y}\); 
      \(l = \min\{\lceil 1/\epsilon\rceil, d_x, d_y\}\): number of columns in each COD sketch;
      \(N\): length of the sliding window;
      \(\theta = \epsilon N\): snapshot registration threshold.
    }
    
    \(\hat{\boldsymbol{A}} \leftarrow \boldsymbol{0}^{d_x\times l},\quad \hat{\boldsymbol{B}} \leftarrow \boldsymbol{0}^{d_y\times l}\).\;
    \(S \leftarrow\) an empty queue.\;
    
    \(\hat{\boldsymbol{A}}' \leftarrow \boldsymbol{0}^{d_x\times l},\quad \hat{\boldsymbol{B}}' \leftarrow \boldsymbol{0}^{d_y\times l}\).\;
    \(S' \leftarrow\) an empty queue.\;
\end{algorithm}

In this section, we introduce our method, \oursolution (\underline{S}pace \underline{O}ptimal \underline{C}o-\underline{O}ccurring \underline{D}irections). As we mentioned earlier, our algorithm is inspired by the \(\lambda\)-snapshot method for \(\epsilon\)-approximation frequency estimation over sliding windows~\cite{LeeT06}.
The key idea is simple yet powerful: as new data arrives, we continuously monitor the dominant (i.e., most significant) co-occurring direction. When the product of the norms of the corresponding vectors exceeds a preset threshold \(\theta\), we “register” a snapshot capturing that direction and remove it from the sketch. This selective registration guarantees that only directions with substantial contributions are maintained—ensuring both accuracy and space efficiency.

 In this section, we focus on the \emph{normalized sliding window} case, where every incoming vector is unit-norm (\(\|\boldsymbol{x}_i\| = \|\boldsymbol{y}_i\| = 1\)). In this case, we have \(\|\boldsymbol{A}_W\|_F^2 = \|\boldsymbol{B}_W\|_F^2 = N\) (with \(N\) being the window length). In Section \ref{sec:unnormalized-setting}, we extend the method to handle the general case (\(R\geq 1\)), referred to as the unnormalized sliding window.

\begin{algorithm}[t]
    \caption{\oursolution: Update(\(\boldsymbol{x}_i,\boldsymbol{y}_i\))}
    \label{alg:simple-update}
    \DontPrintSemicolon
    \KwInput{
     $\boldsymbol{x}_i$: the column vector of $\boldsymbol{X}$ arriving at timestamp $i$;
    $\boldsymbol{y}_i$: the column vector of $\boldsymbol{Y}$ arriving at timestamp $i$
    }
    
    \If{\(i \equiv 1 \pmod{N}\)} 
    {        
        \((\hat{\boldsymbol{A}}, \hat{\boldsymbol{B}}) \leftarrow (\hat{\boldsymbol{A}}', \hat{\boldsymbol{B}}')\).
    
        \((\hat{\boldsymbol{A}}', \hat{\boldsymbol{B}}') \leftarrow (\boldsymbol{0}^{d_x\times l}, \boldsymbol{0}^{d_y\times l})\).

        \(S \leftarrow S'\) and reinitialize \(S'\) as empty.
    }
    
    \While{\(S[0].t + N \le i\)}
    {
        \(S\).POPLEFT() \tcp*[r]{\small Remove expired snapshots}
    }
    
    \(\hat{\boldsymbol{A}},\hat{\boldsymbol{B}} \leftarrow \text{COD}(\hat{\boldsymbol{A}},\hat{\boldsymbol{B}},\boldsymbol{x}_i,\boldsymbol{y}_i)\).\;
    \While{\(\|\hat{\boldsymbol{a}}_1\|_2 \cdot \|\hat{\boldsymbol{b}}_1\|_2 \ge \theta\)}
    {
        Append a snapshot to \(S\): record
        \(
        (\boldsymbol{v}=\hat{\boldsymbol{a}}_1,\, \boldsymbol{u}=\hat{\boldsymbol{b}}_1,\, s=S[-1].t,\, t=i).
        \)
        
        Remove the first row from both \(\hat{\boldsymbol{A}}\) and \(\hat{\boldsymbol{B}}\).\;
        $\hat{\boldsymbol{a}}_2, \hat{\boldsymbol{b}}_2$ becomes the new $\hat{\boldsymbol{a}}_1, \hat{\boldsymbol{b}}_1$
    }
    
    Repeat Lines 5-11 for the auxiliary sketch\;
\end{algorithm}

\subsection{Algorithm Description}

\textbf{Sketch Structure.}
To handle the sliding window efficiently, our proposed \oursolution maintains two separate COD sketches:
\begin{itemize}[topsep=0.5mm, partopsep=0pt, itemsep=0pt, leftmargin=10pt] 
  \item A {\em primary sketch} \((\hat{\boldsymbol{A}}, \hat{\boldsymbol{B}})\) along with its snapshot queue \(S\), which summarizes data within the current sliding window.
  \item An {\em auxiliary sketch} \((\hat{\boldsymbol{A}}', \hat{\boldsymbol{B}}')\) with its snapshot queue \(S'\), which accumulates the most recent updates.
\end{itemize}

Algorithm~\ref{alg:initialize} provides the pseudo-code for initializing the two sketches. It sets both the primary and auxiliary sketches to zero matrices and initializes the queue as empty.

\htitle{Update Algorithm.} When a new pair \((\boldsymbol{x}_i,\boldsymbol{y}_i)\) arrives, we update both sketches using the COD routine. The update procedure, detailed in Alg.\ \ref{alg:simple-update}, involves three main steps:

  {\em (1) Window refresh} (Lines 1-4): Every \(N\) timestamps, the auxiliary sketch and its snapshots replaces that of the primary sketch to discard expired data outside the current sliding window. This swap mitigates the accumulated influence of outdated data outside the sliding window; see the theoretical analysis for details.
  
  {\em (2) Snapshot expiration} (Lines 5-6): It removes any snapshots from both the primary queue \(S\) and the auxiliary queue \(S'\) that have fallen outside the window.
  
  {\em (3) Registering dominant directions} (Lines 7-12): After updating a sketch, we check if the product of the norms of its first (i.e., most significant) column vectors exceeds \(\theta\). When 
  \(
  \|\hat{\boldsymbol{a}}_1\|_2 \cdot \|\hat{\boldsymbol{b}}_1\|_2 \ge \theta,
  \)
  a snapshot capturing these vectors (and the current timestamp information) is appended to the appropriate queue, and the corresponding column is removed from the sketch. This is the core mechanism of our method: only when a co-occurring direction is sufficiently significant do we register it.

\htitle{Complexity Analysis.} Next, we analyze the running cost for \oursolution. For each incoming pair $(\boldsymbol{x}_i,\boldsymbol{y}_i)$, we need to feed them to the COD algorithm (Line 7 of Alg.\ \ref{alg:simple-update}). It needs to perform Lines 6-12 in Alg.\ \ref{alg:cod} every time to obtain the singular values of $\boldsymbol{R}_X\boldsymbol{R}_Y^T$. Performing QR decomposition on $\hat{\boldsymbol{A}}$ and $\hat{\boldsymbol{B}}$ takes $O(d_xl^2)$ and $O(d_yl^2)$ time, respectively. And then computing $\boldsymbol{R}_X\boldsymbol{R}_Y^T$ and performing SVD on matrix $\boldsymbol{R}_X\boldsymbol{R}_Y^T$ cost $O(l^3)$ time. Thereby, the computational bottleneck of the algorithm is the QR decomposition performed within each COD update on the \(d_x \times l\) and \(d_y \times l\) matrices. Hence, each update runs in \(O((d_x+d_y)l^2)\) time.

For space cost, by setting the sketch size $l = min(\lceil \frac{1}{\epsilon}\rceil,d_x,d_y)$, the memory cost of the two COD sketches is bounded by $O(\frac{d_x+d_y}{\epsilon})$. Additionally, setting the register threshold $\theta = \epsilon N$ results in the number of snapshots being $O(\frac{1}{\epsilon})$. In total, the space cost for the whole DS-COD sketch is $O(\frac{d_x+d_y}{\epsilon})$. For the choice of $l$ and $\theta$, we will include more discussions in our later analysis.

\subsection{A Fast Update Algorithm}

A major computational bottleneck in the basic \oursolution update (see Alg.\ \ref{alg:simple-update}) is the repeated full matrix decomposition, which costs \(O((d_x+d_y)l^2)\) per update. In fact, the original COD method defers a full decomposition until the sketch accumulates \(l\) columns, resulting in an amortized cost of \(O((d_x+d_y)l)\) per column. However, in our sliding window scenario, we must process every column to ensure that no significant direction is missed.

We observe that the COD analysis only requires the maintained matrices \(\boldsymbol{Q}_X\) and \(\boldsymbol{Q}_Y\) to be orthonormal; the specific upper triangular structure of \(\boldsymbol{R}_X\) and \(\boldsymbol{R}_Y\) is not essential. Motivated by this observation, we aim to incrementally maintain the decomposition by ensuring that \(\boldsymbol{Q}_X\) and \(\boldsymbol{Q}_Y\) remain orthonormal without recomputing a full decomposition from scratch each time. This leads to our novel \emph{FastUpdate Algorithm} (Alg.\ \ref{alg:fast-update}), which avoids unnecessary full decompositions by updating the existing orthonormal basis incrementally, in a manner akin to the Gram–Schmidt process.

To design this strategy, we additionally maintain four matrices:
\(
\boldsymbol{Q}_X,\ \boldsymbol{R}_X\) and \(\boldsymbol{Q}_Y,\ \boldsymbol{R}_Y,
\)
which capture the decomposition of the primary sketches \(\hat{\boldsymbol{A}}\) and \(\hat{\boldsymbol{B}}\), respectively.

\begin{algorithm}[t]
    \caption{IncDec: Incremental Decomposition}
    \label{alg:ID}
    \DontPrintSemicolon
    \KwInput{\(\boldsymbol{Q}\in \mathbb{R}^{d\times l}\), \(\boldsymbol{R}\in \mathbb{R}^{l\times l}\), and \(\boldsymbol{x}\in \mathbb{R}^d\)}
    \KwOutput{\(\boldsymbol{Q}'\in \mathbb{R}^{d\times (l+1)}\), \(\boldsymbol{R}'\in \mathbb{R}^{(l+1)\times (l+1)}\)}
    
    \(
    \boldsymbol{x}' \leftarrow \boldsymbol{x} - \sum_{i=1}^l \langle \boldsymbol{q}_i,\boldsymbol{x} \rangle\, \boldsymbol{q}_i.
    \)
    
    \tcp{\small It is easy to verify that vector $\boldsymbol{x}'/\left\| \boldsymbol{x}'\right\|$ is a unit vector orthogonal to every column vector of $\boldsymbol{Q}$}
    % The vector \(\boldsymbol{x}'/\|\boldsymbol{x}'\|_2\) is unit norm and orthogonal to every column of \(\boldsymbol{Q}\).

    \(
    \boldsymbol{v} \leftarrow \Bigl(\langle \boldsymbol{q}_1,\boldsymbol{x} \rangle,\, \langle \boldsymbol{q}_2,\boldsymbol{x} \rangle,\, \cdots,\, \langle \boldsymbol{q}_l,\boldsymbol{x} \rangle\Bigr)^T.
    \)

    \(
    \boldsymbol{Q}' \leftarrow \begin{bmatrix}
        \boldsymbol{Q} & \boldsymbol{x}'/\|\boldsymbol{x}'\|_2
    \end{bmatrix}.
    \)

    \(
    \boldsymbol{R}' \leftarrow \begin{bmatrix}
        \boldsymbol{R} & \boldsymbol{v} \\
        \boldsymbol{0} & \|\boldsymbol{x}'\|_2
    \end{bmatrix}.
    \)
    
    \Return \(\boldsymbol{Q}', \boldsymbol{R}'\).
\end{algorithm}

\htitle{Incremental Decomposition.}  When a new vector \(\boldsymbol{x}\) (or \(\boldsymbol{y}\)) arrives, we update the current decomposition using the procedure described in Alg.\ \ref{alg:ID} (IncDec). Briefly, we compute the residual:
\(
\boldsymbol{x}' = \boldsymbol{x} - \sum_{i=1}^l \langle \boldsymbol{q}_i,\boldsymbol{x} \rangle\, \boldsymbol{q}_i,
\)
where \(\{\boldsymbol{q}_i\}_{i=1}^l\) are the columns of the current basis \(\boldsymbol{Q}\) (see Alg.\ \ref{alg:ID}, Line~1). Normalizing \(\boldsymbol{x}'\) produces a new unit vector orthogonal to the existing basis, and the inner products \(\langle \boldsymbol{q}_i,\boldsymbol{x} \rangle\) form a vector \(\boldsymbol{v}\) that updates the upper triangular matrix \(\boldsymbol{R}\) (Lines~2--3 of Alg.\ \ref{alg:ID}). This incremental update requires only \(O(dl)\) time per update, which is dramatically lower than a full decomposition.

\htitle{Fast Update Algorithm.} With the incremental decomposition procedure in place, we now present the complete details of our FastUpdate Algorithm in Alg.\ \ref{alg:fast-update}. The first two steps—window refresh and snapshot expiration—are identical to those in Alg.\ \ref{alg:simple-update}. For the update step, we first invoke Alg.\ \ref{alg:ID} to incrementally update and obtain the decompositions \(\boldsymbol{Q}_X\), \(\boldsymbol{R}_X\), \(\boldsymbol{Q}_Y\), \(\boldsymbol{R}_Y\) (Lines~9--10 in Alg.\ \ref{alg:fast-update}).

\begin{algorithm}[t]
    \caption{\oursolution:FastUpdate($\boldsymbol{x}_i,\boldsymbol{y}_i$)}
    \label{alg:fast-update}
    \DontPrintSemicolon
    % \KwInput{Graph $\mathcal{G}=(\mathcal{V},\mathcal{E})$, feature matrix $\vect{X}$, threshold $\theta$, stopping probability $\alpha$, propagation step $\mathcal{K}$, training budget $\mathcal{B}$,}
    \KwInput{$\boldsymbol{x}_i$: the column vector of $\boldsymbol{X}$ arriving at timestamp $i$;
    $\boldsymbol{y}_i$: the column vector of $\boldsymbol{Y}$ arriving at timestamp $i$}
    
    \While{$S[0].t+N\leq i$} 
    {
    % \Comment{Remove the overdue snapshot}
        $S$.POPLEFT()
    }
    
    $(\hat{\boldsymbol{A}},\hat{\boldsymbol{B}})\leftarrow (\begin{bmatrix} \hat{\boldsymbol{A}} , \boldsymbol{x}_i \end{bmatrix}, \begin{bmatrix} \hat{\boldsymbol{B}} , \boldsymbol{y}_i \end{bmatrix})$

    \If{columns of $\hat{\boldsymbol{A}} \geq l$}
    {
        $(\hat{\boldsymbol{A}},\hat{\boldsymbol{B}}) \leftarrow COD(\hat{\boldsymbol{A}},\hat{\boldsymbol{B}})$

        $(\boldsymbol{Q}_X,\boldsymbol{R}_X) \leftarrow QR(\hat{\boldsymbol{A}})$
        
        $(\boldsymbol{Q}_Y,\boldsymbol{R}_Y) \leftarrow QR(\hat{\boldsymbol{B}})$
    }
    \Else
    {
        $(\boldsymbol{Q}_X,\boldsymbol{R}_X)\leftarrow IncDec(\boldsymbol{Q}_X,\boldsymbol{R}_X,\boldsymbol{x}_i)$
        
        $(\boldsymbol{Q}_Y,\boldsymbol{R}_Y)\leftarrow IncDec(\boldsymbol{Q}_Y,\boldsymbol{R}_Y,\boldsymbol{y}_i)$
    }

    $(\boldsymbol{U},\boldsymbol{\Sigma},\boldsymbol{V}) \leftarrow SVD(\boldsymbol{R}_X\boldsymbol{R}_Y^T)$
    
    \For{$i = 1,\cdots,l$}
    {
        \If{$\sigma_i\geq \theta$}
        {
            $S$ append snapshot ($\boldsymbol{v}=\boldsymbol{Q}_X\boldsymbol{u}_i\sqrt{\sigma_i}$, $\boldsymbol{u} = \boldsymbol{Q}_Y\boldsymbol{v}_i\sqrt{\sigma_i}$, $s=S[-1].t$, $t=i$)

            $\hat{\boldsymbol{A}}{'}\leftarrow \hat{\boldsymbol{A}} - \boldsymbol{Q}_X\boldsymbol{u}_i\boldsymbol{u}_i^T\boldsymbol{R}_X$

            $\hat{\boldsymbol{B}}{'} \leftarrow \hat{\boldsymbol{B}} - \boldsymbol{Q}_Y\boldsymbol{v}_i\boldsymbol{v}_i^T\boldsymbol{R}_Y$
            
            $\boldsymbol{R}_X' \leftarrow \boldsymbol{R}_X - \boldsymbol{u}_i\boldsymbol{u}_i^T\boldsymbol{R}_X$

            $\boldsymbol{R}_Y' \leftarrow \boldsymbol{R}_Y - \boldsymbol{v}_i\boldsymbol{v}_i^T\boldsymbol{R}_Y$

            $(\hat{\boldsymbol{A}},\hat{\boldsymbol{B}},\boldsymbol{R}_X,\boldsymbol{R}_Y) \leftarrow (\hat{\boldsymbol{A}}{'},\hat{\boldsymbol{B}}{'},\boldsymbol{R}_X',\boldsymbol{R}_Y')$
        }
    }
    
Repeat the above process to the auxiliary sketch  
    
\end{algorithm}

Once these updated decompositions are available, we compute the product \(\boldsymbol{R}_X\boldsymbol{R}_Y^T\) and perform a singular value decomposition (SVD) on it (which takes \(O(l^3)\) time, as shown in Line~11). The SVD yields singular values \(\{\sigma_i\}\) and the associated singular vectors. We then check whether any singular value \(\sigma_i\) meets or exceeds the snapshot threshold \(\theta\). For each \(\sigma_i \ge \theta\), a snapshot is generated immediately by computing:
\(
\hat{\boldsymbol{a}}_i = \boldsymbol{Q}_X\,\boldsymbol{u}_i\sqrt{\sigma_i}\) and  \(\hat{\boldsymbol{b}}_i = \boldsymbol{Q}_Y\,\boldsymbol{v}_i\sqrt{\sigma_i},
\)
where \(\boldsymbol{u}_i\) and \(\boldsymbol{v}_i\) are the singular vectors corresponding to \(\sigma_i\), and it holds that
\(
\|\hat{\boldsymbol{a}}_i\|_2 \,\|\hat{\boldsymbol{b}}_i\|_2 = \sigma_i.
\)
This snapshot computation requires \(O((d_x+d_y)l)\) time. Note that our snapshot registration process differs from that in Alg.\ \ref{alg:cod}. Here, we must update \(\hat{\boldsymbol{A}}\), \(\hat{\boldsymbol{B}}\), \(\boldsymbol{R}_X\), and \(\boldsymbol{R}_Y\) so that the invariant
\(
\hat{\boldsymbol{A}} = \boldsymbol{Q}_X\boldsymbol{R}_X\) and \(\hat{\boldsymbol{B}} = \boldsymbol{Q}_Y\boldsymbol{R}_Y
\)
remains preserved after the update. Although this update strategy is different, we will show that it still produces the correct result. Finally, as detailed in Lines~15–19 of Alg.\ \ref{alg:fast-update}, the corresponding snapshot vectors are removed from \(\hat{\boldsymbol{A}}\), \(\hat{\boldsymbol{B}}\), \(\boldsymbol{R}_X\), and \(\boldsymbol{R}_Y\) using our incremental decomposition results; the correctness of these removals is supported by Lemma~\ref{lem:fast-equivalent}. Finally, we apply the same update strategy for the auxiliary sketch (Alg. \ref{alg:fast-update} Line 20).  

By replacing full decompositions with an efficient incremental update strategy, our FastUpdate algorithm achieves an amortized update cost of \(O((d_x+d_y)l + l^3)\) per column. This novel improvement retains key properties required for COD while substantially reducing computational overhead, making our approach particularly well-suited for high-dimensional streaming data.

\begin{lemma} \label{lem:fast-equivalent}
    if $\hat{\boldsymbol{A}}=\boldsymbol{Q}_X \boldsymbol{R}_X$, $\hat{\boldsymbol{B}}=\boldsymbol{Q}_Y\boldsymbol{R}_Y$, $(\boldsymbol{U},\boldsymbol{\Sigma},\boldsymbol{V}) = SVD(\boldsymbol{R}_X\boldsymbol{R}_Y^T)$, $\hat{\boldsymbol{A}}{'}= \hat{\boldsymbol{A}} - \boldsymbol{Q}_X\boldsymbol{u}_i\boldsymbol{u}_i^T\boldsymbol{R}_X$, $\hat{\boldsymbol{B}}{'} = \hat{\boldsymbol{B}} - \boldsymbol{Q}_Y\boldsymbol{v}_i\boldsymbol{v}_i^T\boldsymbol{R}_Y$, $\boldsymbol{R}_X' = \boldsymbol{R}_X - \boldsymbol{u}_i\boldsymbol{u}_i^T\boldsymbol{R}_X$ and $\boldsymbol{R}_Y' = \boldsymbol{R}_Y - \boldsymbol{v}_i\boldsymbol{v}_i^T\boldsymbol{R}_Y$, then $(\boldsymbol{Q}_X,\boldsymbol{R}_X')$ is an orthogonal decomposition of $\hat{\boldsymbol{A}}{'}$, $(\boldsymbol{Q}_Y,\boldsymbol{R}_Y')$ is the orthogonal decomposition of $\hat{\boldsymbol{B}}{'}$ and $(\hat{\boldsymbol{A}}{'}, \hat{\boldsymbol{B}}{'})$ is the same as removing $(\hat{a_i},\hat{b_i})$ from $(\hat{\boldsymbol{A}},\hat{\boldsymbol{B}})$, that is, $\hat{\boldsymbol{A}}{'}\hat{\boldsymbol{B}}{'}^T = \hat{\boldsymbol{A}}\hat{\boldsymbol{B}}^T - \hat{a_i}\hat{b_i}^T$.
\end{lemma}

All omitted proofs can be found in Appendix \ref{app:proofs}.

\htitle{Query algorithm.}
The query procedure for \oursolution\ operate through a direct aggregation of the COD sketch matrices and snapshots from the primary sketch. Specifically, the current sketches $(\hat{\boldsymbol{A}},\hat{\boldsymbol{B}})$ are horizontally concatenated with matrices $(\hat{\boldsymbol{C}},\hat{\boldsymbol{D}})$ stacked by the non-expiring snapshots, constructing an augmented representation that preserves historical data integrity within the window while incorporating current COD sketches.

\begin{algorithm}[t]
    \caption{\oursolution:Query()}
    \label{alg:query-normalized}
    \DontPrintSemicolon
    \KwOutput{$\boldsymbol{A}_{aug}$ and $\boldsymbol{B}_{aug}$}

    $\boldsymbol{A}_{aug} = 
    \begin{bmatrix}
     \hat{\boldsymbol{A}},\boldsymbol{C}   
    \end{bmatrix}$, where $\boldsymbol{C}$ is stacked $s_i.\boldsymbol{u}$ for all $s_i \in S$
    
    $\boldsymbol{B}_{aug} = 
    \begin{bmatrix}
     \hat{\boldsymbol{B}},\boldsymbol{D}   
    \end{bmatrix}$, where $\boldsymbol{D}$ is stacked $s_i.\boldsymbol{v}$ for all $s_i \in S$

    \Return $\boldsymbol{A}_{aug}, \boldsymbol{B}_{aug}$
    
\end{algorithm}

\htitle{Theoretical Analysis.} To compute $\boldsymbol{Q}_X\boldsymbol{u}_i\boldsymbol{u}_i^T\boldsymbol{R}_X$, we should first compute $\boldsymbol{Q}_X\boldsymbol{u}_i$ in $O(d_xl)$ time, $\boldsymbol{u}_i^T\boldsymbol{R}_X$ in $O(l^2)$ time and then multiply $\boldsymbol{Q}_X\boldsymbol{u}_i$ by $\boldsymbol{u}_i^T\boldsymbol{R}_X$ in $O(d_xl)$ time to obtain $\boldsymbol{Q}_X\boldsymbol{u}_i\boldsymbol{u}_i^T\boldsymbol{R}_X$. Similarly, we can compute $\boldsymbol{Q}_Y\boldsymbol{v}_i\boldsymbol{v}_i^T\boldsymbol{R}_Y$ in $O(d_yl)$ time. For $\boldsymbol{u}_i\boldsymbol{u}_i^T\boldsymbol{R}_X$, we can first calculate $\boldsymbol{u}_i^T\boldsymbol{R}_X$ in $O(l^2)$ time and then multiply $\boldsymbol{u}_i$ by $\boldsymbol{u}_i^T\boldsymbol{R}_X$ to attain $\boldsymbol{u}_i\boldsymbol{u}_i^T\boldsymbol{R}_X$ in $O(l^2)$ time. We can also compute $\boldsymbol{v}_i\boldsymbol{v}_i^T\boldsymbol{R}_Y$ in $O(l^2)$ time. As a result, extracting a snapshot from $(\hat{\boldsymbol{A}},\hat{\boldsymbol{B}})$ takes $O((d_x+d_y)l)$ time. However, it is noteworthy that after executing Lines 17 and 18, $\boldsymbol{R}_X$ and $\boldsymbol{R}_Y$ are not necessarily upper triangular matrices. What can be guaranteed is that $(\boldsymbol{Q}_X,\boldsymbol{R}_X')$ and $(\boldsymbol{Q}_Y,\boldsymbol{R}_Y')$ are the orthogonal column decomposition of $\hat{\boldsymbol{A}}$ and $\hat{\boldsymbol{B}}$ respectively. However, as we mentioned earlier, a closer examination of the COD algorithm's proof in \cite{MrouehMG17} reveals that the upper triangular structure of $\boldsymbol{R}_X$ and $\boldsymbol{R}_Y$ is nonessential to the algorithm's validity. The algorithm's correctness hinges solely on $\boldsymbol{Q}_X$ and $\boldsymbol{Q}_Y$ being column orthogonal. Notably, our proposed Algorithm \ref{alg:ID} is designed to handle decompositions with non-triangular $\boldsymbol{R}_X$ and $\boldsymbol{R}_Y$. It remains valid for the pairs $(\boldsymbol{Q}_X,\boldsymbol{R}_X')$ and $(\boldsymbol{Q}_Y,\boldsymbol{R}_Y')$ as long as $\boldsymbol{Q}_X$ and $\boldsymbol{Q}_Y$ are column orthogonal. Consequently, decompositions $(\boldsymbol{Q}_X,\boldsymbol{R}_X')$ and $(\boldsymbol{Q}_Y,\boldsymbol{R}_Y')$ remain valid for the COD algorithm and Algorithm \ref{alg:ID}, regardless of the upper triangular structure of $\boldsymbol{R}_X$ and $\boldsymbol{R}_Y$.

Suppose that there exist $m$ singular values surpassing the threshold $\theta$ in Line 13 from the beginning to current timestamp $t$, aligned as $\sigma_1\geq \sigma_2 \geq \sigma_3 \geq \cdots \geq \sigma_m \geq \theta$, extracting these $m$ snapshots takes $O(m(d_x+d_y)l)$ time in total. Since $m\theta \leq \sum_{i=1}^m\sigma_i \leq \left\| \boldsymbol{A}_{1,t}\boldsymbol{B}_{1,t}^T\right\|_{*} \leq \sum_{i=1}^t \left\|\boldsymbol{x}_i\boldsymbol{y}_i^T\right\|_{*} \leq t$, we have $m \leq t/\theta$. As a result, it takes $O(\frac{t(d_x+d_y)l}{\theta})$ over $t$ timestamps, and the amortized time per column is $O(\frac{(d_x+d_y)l}{\theta})$. With $\theta = \epsilon N = N / l$, this amortized time is $O((d_x+d_y)l^2/N)$. Assuming $N = \Omega(l)$, this amortization for a single update remains $O((d_x+d_y)l + l^3)$, which is a rational assumption. Thus, compared to Alg.\ \ref{alg:simple-update} costing $O((d_x+d_y)l^2)$ for each update, Alg.\ \ref{alg:fast-update} only takes amortized $O((d_x+d_y)l + l^3)$ time for each update, which is faster.

Finally, we present the following theorem about the space cost, update cost and error guarantee of our \oursolution.

\begin{theorem}\label{thm:socod-normalized}
Let \(\{(\boldsymbol{x}_t,\boldsymbol{y}_t)\}_{t\ge1}\) be a stream of data with \(\|\boldsymbol{x}_t\|=\|\boldsymbol{y}_t\|=1\) for all \(t\), and let \(\boldsymbol{X}_W\) and \(\boldsymbol{Y}_W\) denote the sliding window matrices as defined in Definition~\ref{def:amm}. Given a window size \(N\) and relative error \(\epsilon\), the \oursolution algorithm outputs matrices \(\boldsymbol{A}_{aug}\in\mathbb{R}^{d_x\times O(\frac{1}{\epsilon})}\) and \(\boldsymbol{B}_{aug}\in\mathbb{R}^{d_y\times O(\frac{1}{\epsilon})}\) such that, if the register threshold is set to \(\theta = \epsilon N\) and 
\(
l = \min\Bigl(\Bigl\lceil \frac{1}{\epsilon}\Bigr\rceil,\, d_x,\, d_y\Bigr),
\)
then
\[
\Bigl\|\boldsymbol{X}_W\boldsymbol{Y}_W^\top - \boldsymbol{A}_{aug}\boldsymbol{B}_{aug}^\top\Bigr\|_2 \le 8\epsilon N = 8\epsilon\,\|\boldsymbol{X}_W\|_F\,\|\boldsymbol{Y}_W\|_F.
\]
Furthermore, the \oursolution sketch uses \(O\Bigl(\frac{d_x+d_y}{\epsilon}\Bigr)\) space and supports each update in \(O((d_x+d_y)l+l^3)\) time with Alg.\ \ref{alg:fast-update}.
\end{theorem}

\section{General Unnormalized Model}\label{sec:unnormalized-setting}

\begin{algorithm}[t]
    \caption{\newsolution:Initialize($d_x,d_y,l,N,R,\epsilon$)}
    \label{alg:initialize-ml}
    \DontPrintSemicolon
    \KwInput{The dimension of $\boldsymbol{X}$ and $\boldsymbol{Y}$: $d_x$ and $d_y$ respectively, number of columns in the \oursolution sketch $l = min(\lceil \frac{1}{\epsilon}\rceil,d_x,d_y)$, the length of sliding window $N$, upper bound of squared norms: $R$, error parameter $\epsilon$}

    $L \leftarrow \lceil log_2{R} \rceil$; $M \leftarrow $ an empty list

    \For{$i = 0,\cdots,L-1$}
    {
        $M.\textit{append}(\text{\oursolution}.\textit{Initialize}(d_x,d_y,l,N,2^i\epsilon N))$
    }
    
\end{algorithm}

In this section, we extend \oursolution to handle unnormalized data, where the squared norms of the input vectors satisfy \( \|\boldsymbol{x}_i\|_2^2,\;\|\boldsymbol{y}_i\|_2^2 \in [1,R] \). Following the approach in \cite{LeeT06} for mutable sliding window sizes, we construct a multi-layer extension of \oursolution, which we denote by \newsolution (Multi-Layer SO-COD). In this framework, the data stream is processed through \( L = \lceil \log_2 R \rceil \) layers. Each layer \( i \) (for \( i = 0,1,\dots,L-1 \)) maintains an independent \oursolution sketch with a distinct register threshold \( \theta_i \) defined by \( \theta_i = 2^i\epsilon N \). That is, in layer \( i \), a snapshot is registered when \( \|\hat{\boldsymbol{a}}_i\|_2 \,\|\hat{\boldsymbol{b}}_i\|_2 \ge 2^i\epsilon N \). By constraining the number of snapshots per layer to \( O(1/\epsilon) \), the overall space complexity of \newsolution is \( O((d_x+d_y)/\epsilon \cdot \log_2 R) \).

\subsection{Algorithm Description}
    \htitle{Multi-Layer Sketch Structure.}
Algorithm~\ref{alg:initialize-ml} specifies the initialization of a \newsolution sketch, thereby defining the multi-layer structure. In Line 1, we compute the number of layers \( L = \lceil \log_2 R \rceil \). For each layer \( i \in \{0, 1, \dots, L-1\} \), an independent \oursolution sketch is initialized with the register threshold \( \theta_i = 2^i\epsilon N \) and added to the list $M$. Consequently, in layer \( i \), the \oursolution sketch will generate a snapshot when \( \|\hat{\boldsymbol{a}}_i\|_2 \,\|\hat{\boldsymbol{b}}_i\|_2 \ge 2^i\epsilon N \). To minimize redundancy and retain only the most significant information, we restrict the number of snapshots per layer to \( O(1/\epsilon) \). As a result, the total space required by \newsolution is \( O((d_x+d_y)/\epsilon \cdot \log_2 R) \).

    \begin{algorithm}[t]
    \caption{\newsolution:Update($\boldsymbol{x}_i,\boldsymbol{y}_i$)}
    \label{alg:update-ml}
    \DontPrintSemicolon
    \KwInput{$\boldsymbol{x}_i$: the column vector of $\boldsymbol{X}$ arriving at timestamp $i$;
    $\boldsymbol{y}_i$: the column vector of $\boldsymbol{Y}$ arriving at timestamp $i$}

    \For{$j = 0,\cdots,L-1$}
    {
        \While{$len(M[j].S)> \lsn$ or $M[j].S[0].t \leq i-N$}
        {
            $M[j].S.POPLEFT()$
        }
        \If{$\left\|\boldsymbol{x}_i\right\|_2\left\|\boldsymbol{y}_i\right\|_2 \geq 2^j\epsilon N$}
        {
            $M[j].S$ append snapshot ($\boldsymbol{u}=\boldsymbol{x}_i$, $\boldsymbol{v}=\boldsymbol{y}_i$, $s=M[j].S[-1].t$, $t = i$)
            
            $M[j].S'$  append snapshot ($\boldsymbol{u} = \boldsymbol{x}_i$, $\boldsymbol{v} = \boldsymbol{y}_i$, $s=M[j].S'[-1].t$, $t = i$)
        }
        \Else
        {
            $M[j].\textit{FastUpdate}(\boldsymbol{x}_i,\boldsymbol{y}_i)$
        }
    }
    
\end{algorithm}

\htitle{Update Algorithm.} As shown in Alg.\ \ref{alg:update-ml}, each incoming column pair \( (\boldsymbol{x}_i,\boldsymbol{y}_i) \) is processed across all layers. To ensure that the number of snapshots in each layer remains within the prescribed bound of \( O(1/\epsilon) \), we explicitly set this bound to \( \lsn \); that is, the algorithm first checks whether the number of snapshots exceeds \( \lsn \) (and prunes expired snapshots promptly, as indicated in Lines 2--3). Then, for each layer \( j \), if the product of $\xinorm$ and $\yinorm$ exceed the threshold $2^j\epsilon N$, we  generate a snapshot $(u=\boldsymbol{x}_i,v=\boldsymbol{y}_i)$ directly, preserving all information of $(\boldsymbol{x}_i,\boldsymbol{y}_i)$ without introducing approximation error for $\boldsymbol{x}_i\boldsymbol{y}_i^T$ to reduce additional calculation (Lines 4 to 6). Otherwise, the update procedure applies the column pair \( (\boldsymbol{x}_i,\boldsymbol{y}_i) \) to update the corresponding \oursolution sketch \( M[j] \) (Line 8) via the \oursolution.\textit{FastUpdate} procedure.  Due to the direct update mechanism, we can better restrict the times of extracting snapshots from sketch. Specifically, suppose that in layer $j$, there exist $m$ singular values surpassing the threshold $\theta$ in Alg.\ref{alg:fast-update} (Line 13) from the beginning to current timestamp $t$, aligned as $\sigma_1\geq \sigma_2 \geq \sigma_3 \geq \cdots \geq \sigma_m \geq \theta$, extracting these $m$ snapshots takes $O(m(d_x+d_y)l)$ time in total. Since $m\theta \leq \sum_{i=1}^t \left\|\boldsymbol{x}_i\boldsymbol{y}_i^T\right\|_{*} \cdot \mathbb{I}(\xinorm\yinorm <\theta) < t\theta$, where $\mathbb{I}(\xinorm\yinorm <\theta)$ equals $1$ if $\xinorm\yinorm <\theta$ and otherwise $0$. Thereby, we have $m \leq t$, which implies amortized time for extractng the snapshots per layer is $O((d_x+d_y)l)$. As the update is executed in each of the \( L \) layers, the overall time cost per update is \( O(((d_x+d_y)l + l^3)\log R) \).

\htitle{Query Algorithm.} Alg.\ \ref{alg:query} describes the procedure for forming the sketch corresponding to the sliding window \( [t-N+1,t] \). Because of the per-layer constraint on the number of snapshots, a given layer might not contain enough snapshots to cover the entire window and thereby produce a valid sketch. To address this, we select the lowest layer for which the snapshots fully cover the window while minimizing the approximation error. More precisely, a layer is deemed valid if the last expired snapshot before its earliest non-expired snapshot occurs at time \( s \) satisfying \( s \le t-N \). A naive approach would scan all \( O(\log R) \) layers, resulting in a time complexity of \( O(\log R) \); however, since the snapshot density decreases monotonically with increasing layer index (due to the larger register thresholds), a binary search can be employed to reduce the query time complexity to \( O(\log \log R) \).

\htitle{Theoretical Analysis.}
%    \subsection{Algorithm Analysis}
The following theorem demonstrates the error guarantee, space cost and time cost for \newsolution.
\vspace{-1mm}
\begin{theorem}\label{thm:socod-unnormalized}
Let \(\{(\boldsymbol{x}_t,\boldsymbol{y}_t)\}_{t\ge1}\) be a stream of data so that for all \(t\) it holds that
\(
\|\boldsymbol{x}_t\|_2^2,\;\|\boldsymbol{y}_t\|_2^2\in[1,R]
\). Let 
\(
\boldsymbol{X}_W = [\boldsymbol{x}_{t-N+1},\dots,\boldsymbol{x}_t]
\)
and 
\(
\boldsymbol{Y}_W = [\boldsymbol{y}_{t-N+1},\dots,\boldsymbol{y}_t]
\)
denote the sliding window matrices as defined in Def.~\ref{def:amm}. Given window size \(N\) and relative error parameter \(\epsilon\), the \newsolution algorithm outputs matrices
\(
\boldsymbol{A}_{aug}\in\mathbb{R}^{d_x\times O(\frac{1}{\epsilon})}
\)
and 
\(
\boldsymbol{B}_{aug}\in\mathbb{R}^{d_y\times O(\frac{1}{\epsilon})}
\)
such that if the sketch size is set to 
\(
l = \min\Bigl(\Bigl\lceil\frac{1}{\epsilon}\Bigr\rceil,\, d_x,\, d_y\Bigr),
\)
then
\(
\Bigl\|\boldsymbol{X}_W\boldsymbol{Y}_W^\top - \boldsymbol{A}_{aug}\boldsymbol{B}_{aug}^\top\Bigr\|_2 \le 4\epsilon\,\|\boldsymbol{X}_W\|_F\,\|\boldsymbol{Y}_W\|_F.
\)
Furthermore, the \newsolution sketch uses 
\(
O\Bigl(\frac{d_x+d_y}{\epsilon}\log R\Bigr)
\)
space and supports each update in 
\(
O\Bigl(((d_x+d_y)l + l^3)\log R\Bigr)
\)
time.
\end{theorem}

    \begin{algorithm}[t]
    \caption{\newsolution:Query()}
    \label{alg:query}
    \DontPrintSemicolon
    \KwOutput{$\boldsymbol{A}_{aug}$ and $\boldsymbol{B}_{aug}$}

    Find $i = \min_j 1\leq M[j].S[0].s\leq t - N$
    
    $\boldsymbol{A}_{aug} = 
    \begin{bmatrix}
     M[i].\hat{\boldsymbol{A}},\boldsymbol{C}   
    \end{bmatrix}$, where $\boldsymbol{C}$ is stacked $s_j.\boldsymbol{u}$, $\forall$ $s_j \in M[i].S$
    
    $\boldsymbol{B}_{aug} = 
    \begin{bmatrix}
     M[i].\hat{\boldsymbol{B}}, \boldsymbol{D}   
    \end{bmatrix}$, where $\boldsymbol{D}$ is stacked $s_j.\boldsymbol{v}$, $\forall$ $s_j \in M[i].S$

    \Return $\boldsymbol{A}_{aug}, \boldsymbol{B}_{aug}$
    
\end{algorithm}

\label{sec-analysis}

\section{Space Lower Bound}
\label{sec:sec-LowerBound}

We derive a space lower bound for any deterministic algorithm that solves the AMM problem over a sliding window. Prior work in \cite{GhashamiLPW16} established a lower bound for matrix sketching, and \cite{LuoCXY21} extended this result to AMM, as stated in the following lemma.

\begin{lemma}[Lower Bound for AMM \cite{LuoCXY21}]\label{lem:lower-bound-amm}
Consider any AMM algorithm with input matrices \( \boldsymbol{X} \in \mathbb{R}^{d_x \times n} \) and \( \boldsymbol{Y} \in \mathbb{R}^{d_y \times n} \), and outputs \( \boldsymbol{A} \in \mathbb{R}^{d_x \times m} \) and \( \boldsymbol{B} \in \mathbb{R}^{d_y \times m} \) that satisfy
\( \|\boldsymbol{X}\boldsymbol{Y}^T - \boldsymbol{A}\boldsymbol{B}^T\|_2 \le \frac{1}{m}(\|\boldsymbol{X}\|_F \|\boldsymbol{Y}\|_F) \).
Assuming a constant number of bits per word, any such algorithm requires at least \( \Omega((d_x+d_y)m) \) bits of space.
\end{lemma}

We also need the following lemma.

\begin{lemma}[\cite{LuoCXY21}]\label{lem:cardinality}
For any \( \delta > 0,  d_x \le d_y \), there exists a set of matrix pairs 
\( \hat{\mathcal{Z}}_l = \{ (\hat{\boldsymbol{X}}_1,\hat{\boldsymbol{Y}}_1), \dots, (\hat{\boldsymbol{X}}_M,\hat{\boldsymbol{Y}}_M) \} \),
where \( M = 2^{\Omega(l(d_y-l)\log(1/\delta))} \) and each \( \hat{\boldsymbol{X}}_i \in \mathbb{R}^{d_x \times l} \) and \( \hat{\boldsymbol{Y}}_i \in \mathbb{R}^{d_y \times l} \) satisfies 
\( \hat{\boldsymbol{X}}_i^T\hat{\boldsymbol{X}}_i = I_l \) and \( \hat{\boldsymbol{Y}}_i^T\hat{\boldsymbol{Y}}_i = I_l \). Besides, for any \( i \neq j \), 
\( \|\hat{\boldsymbol{X}}_i\hat{\boldsymbol{Y}}_i^T - \hat{\boldsymbol{X}}_j\hat{\boldsymbol{Y}}_j^T\| > \sqrt{2\delta} \).
\end{lemma}

Combining Lemmas~\ref{lem:lower-bound-amm} and \ref{lem:cardinality}, we obtain the following lower bound for the AMM problem over a sliding window.

\begin{theorem}[Lower Bound for AMM over Sliding Window]\label{thm:lower-bound-amm-sw}
    Given $\boldsymbol{X}_W\in\mathbb{R}^{(d_x+1)\times N}, \boldsymbol{Y}_W\in\mathbb{R}^{(d_y+1)\times N}$ and $1\leq \left\|\boldsymbol{x}\right\|_2^2, \left\|\boldsymbol{y}\right\|_2^2\leq R+1$ for all $x\in\boldsymbol{X}_W$ and $y\in\boldsymbol{Y}_W$,
    a deterministic algorithm that returns $(\boldsymbol{A}_W, \boldsymbol{B}_W)$ such that
    \(
    \left\| \boldsymbol{X}_W\boldsymbol{Y}_W^T - \boldsymbol{A}_W\boldsymbol{B}_W^T\right\|_2 \leq \frac{\epsilon}{3}\left\|\boldsymbol{X}_W\right\|_F\left\|\boldsymbol{Y}_W\right\|_F
    \)
where $\epsilon = 1/l$ and $N \geq \frac{1}{2\epsilon}\log{\frac{R}{\epsilon}}$, must use $\Omega(\frac{d_x+d_y}{\epsilon}\log{R})$ space.
\end{theorem}

\section{Experiments}
\label{sec:sec-Experiment}

In this section, we present a comprehensive evaluation of our proposed algorithms against three different baseline algorithms on both synthetic and real-world datasets.

\subsection{Baseline Algorithms}

\begin{figure*}[t]
	\centering
		\begin{tabular}{cccc}
	 			\multicolumn{4}{c}{\hspace{-8mm} \includegraphics[height=2.8mm]{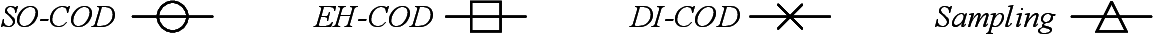}}   \\[-1mm]
                 \includegraphics[height=26mm]{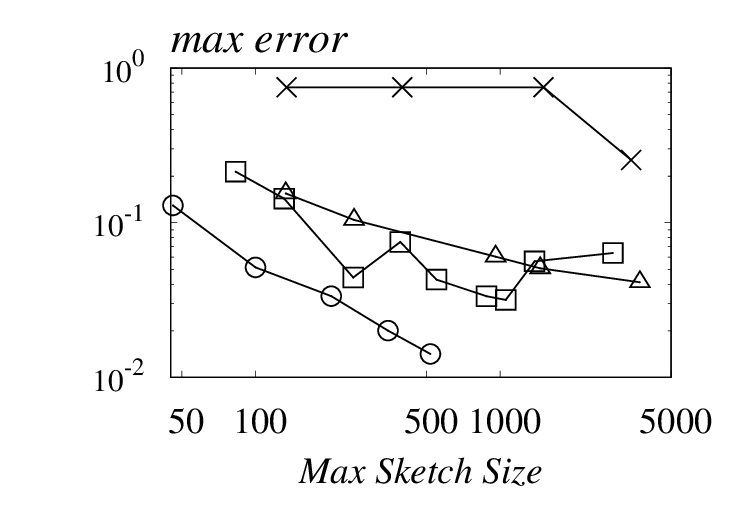} &
			 \includegraphics[height=26mm]{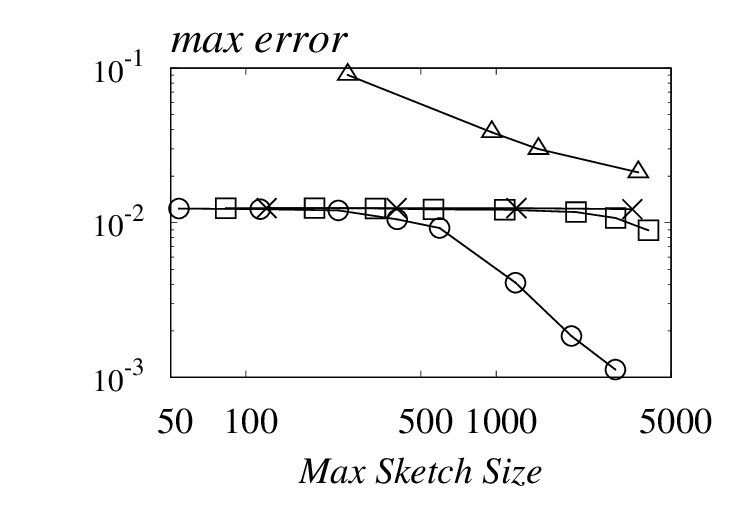} &
			\ \includegraphics[height=26mm]{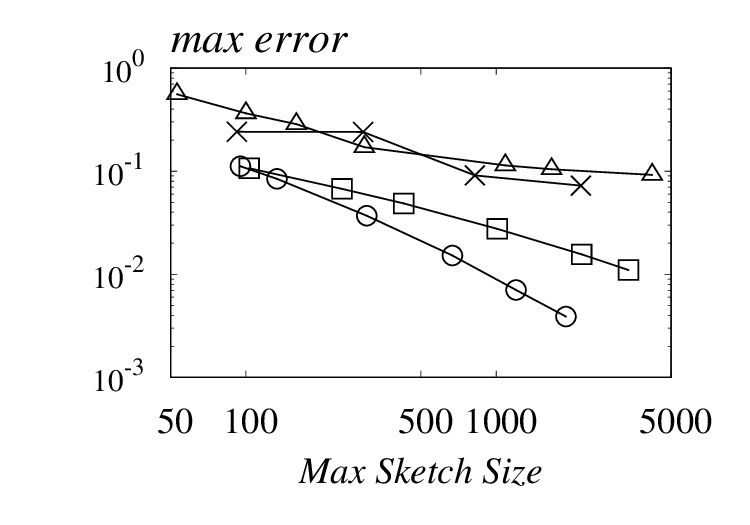} &
			 \includegraphics[height=26mm]{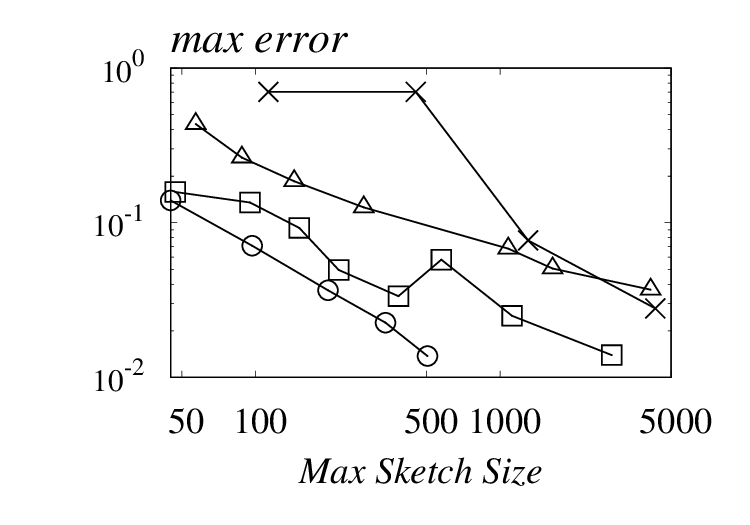}
			\\[-3mm]
             (a) Uniform Random &
                 (b) Random Noisy &
			 (c) Multimodal Data &
			 (d) HPMaX  \\[-1mm]
		\end{tabular}
		\vspace{-3mm}
		\caption{Maximum Sketch Size vs. Maximum Error.} \label{fig:max-error}
		\vspace{-1mm}
\end{figure*}

\begin{figure*}[t]
	\centering
 \vspace{-2mm}
		\begin{tabular}{cccc}
	 	%		\multicolumn{4}{c}{\hspace{-8mm} \includegraphics[height=2.8mm]{figures/socod_legend.pdf}}   \\
			\includegraphics[height=26mm]{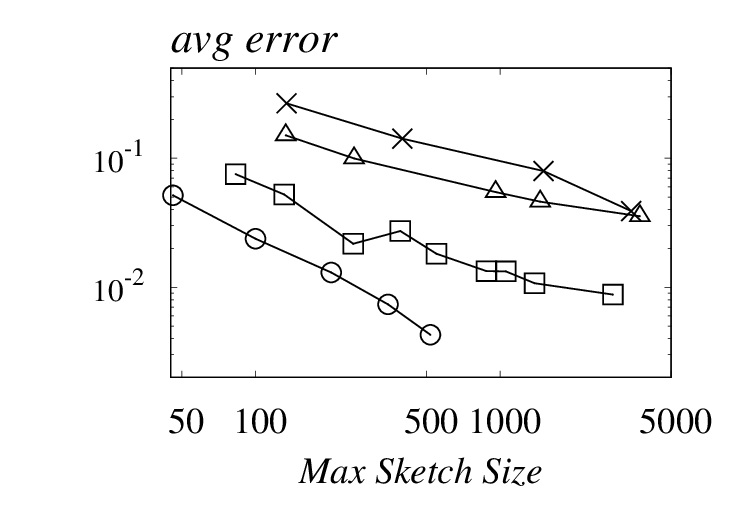} &
             \includegraphics[height=26mm]{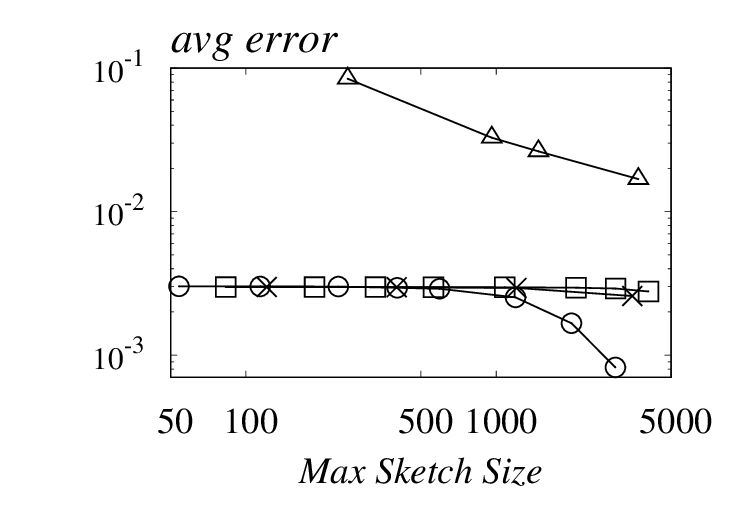} &
			 \includegraphics[height=26mm]{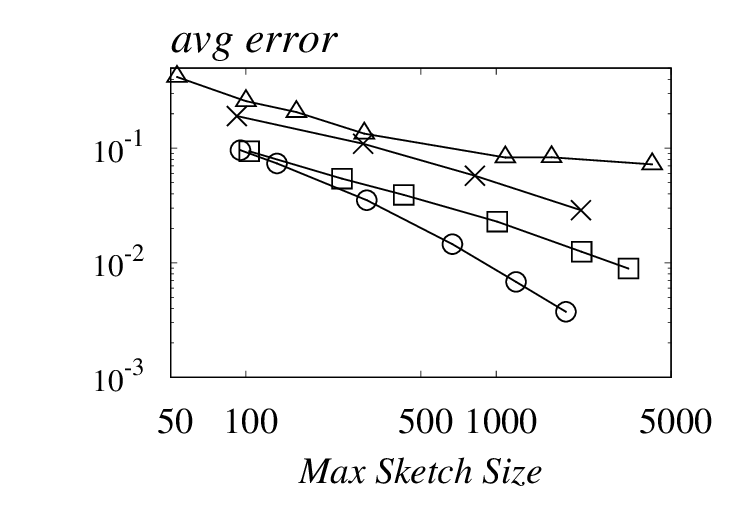} &
			 \includegraphics[height=26mm]{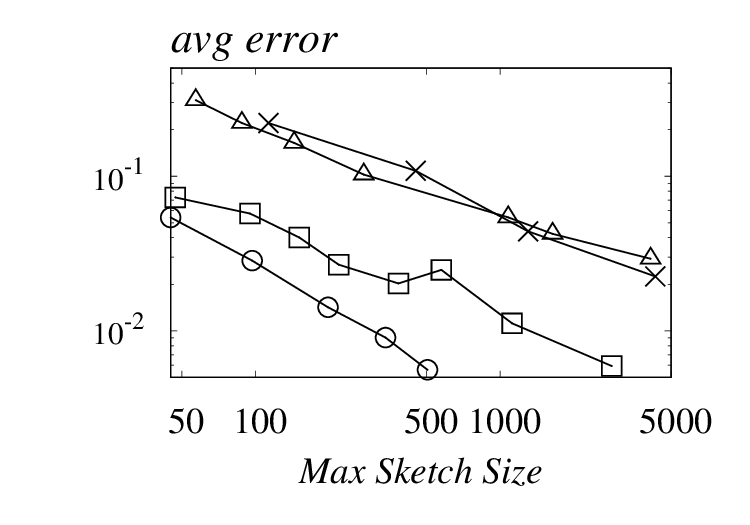}
			\\[-3mm]
                 (a) Uniform Random &
                (b) Random Noisy &
			 (c) Multimodal Data &
			 (d) HPMaX  \\[-1mm]
		\end{tabular}
		\vspace{-3mm}
		\caption{Maximum Sketch Size vs. Average Error.} \label{fig:avg-error}
		% \vspace{-1mm}
\end{figure*}

\begin{figure*}[t]
	\centering
 \vspace{-2mm}
		\begin{tabular}{cccc}
	 	%		\multicolumn{4}{c}{\hspace{-8mm} \includegraphics[height=2.8mm]{figures/socod_legend.pdf}}   \\
			 \includegraphics[height=26mm]{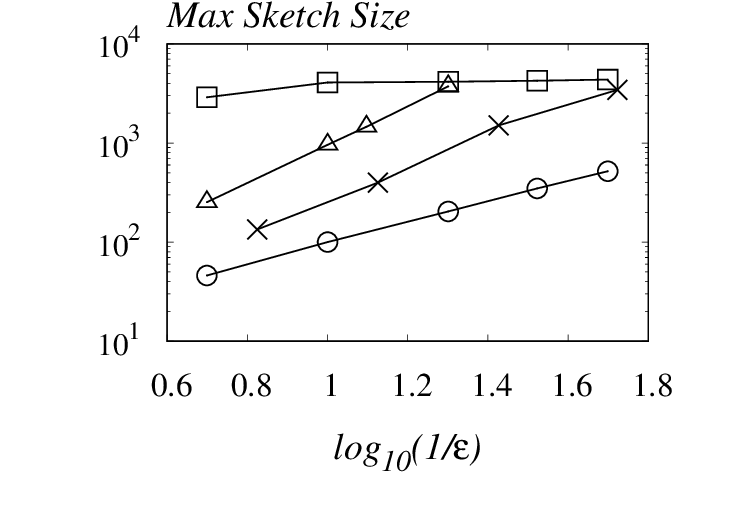} &
            		 \includegraphics[height=26mm]{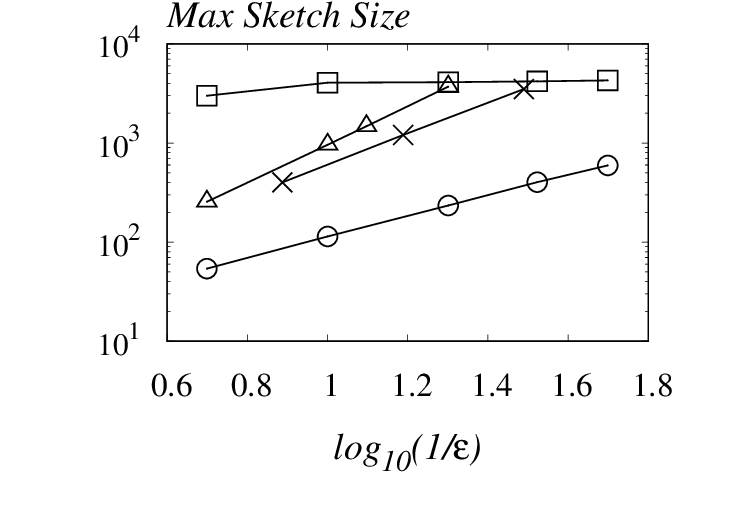} &
			 \includegraphics[height=26mm]{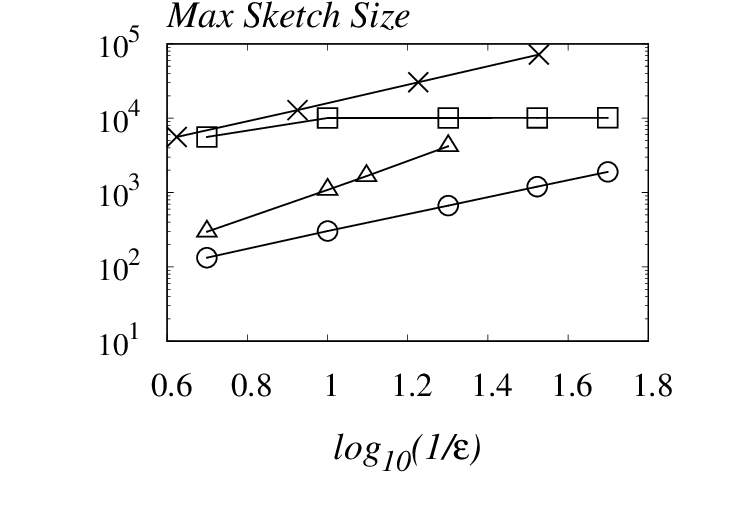} &
			 \includegraphics[height=26mm]{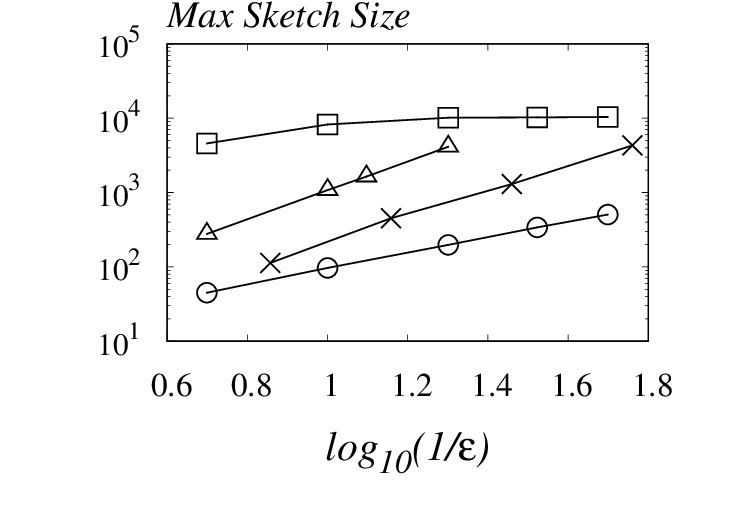}
			\\[-3mm]
             (a) Uniform Random &
                (b) Random Noisy &
			 (c) Multimodal Data &
			(d) HPMaX  \\[-1mm]
		\end{tabular}
		\vspace{-2mm}
		\caption{$\log_{10}(1/\epsilon)$ vs. Maximum Sketch Size ($\log_{10}(\frac{1}{0.25}) \approx 0.6$, \text{and} $\log_{10}(\frac{1}{0.016}) \approx 1.8$).}\label{fig:sketch-size}
		\vspace{-1mm}
\end{figure*}

\htitle{Sampling method.} For the AMM problem, the algorithm samples a small proportion of the matrices. Specifically, each pair of columns $(x_i,y_i)$ is assigned to a priority $\rho=u^{1/(\xinorm\yinorm)}$, where $u$ is uniformly sampled from the interval $(0,1)$ \cite{efraimidis2006weighted}. This priority-based sampling strategy ensures that columns with larger norms are more likely to be selected, thereby preserving the most significant contributions to the matrix product. To achieve an $\epsilon$-approximation guarantee, the algorithm requires $O(\frac{1}{\epsilon^2})$ independent samples selected based on the highest priorities \cite{drineas2006fast, efraimidis2006weighted}. To extend the priority sampling on the sliding window, we use the technique from \cite{babcock2001sampling}, leading to a space complexity of $O(\frac{d_x+d_y}{\epsilon^2}\log{N})$ for the normalized model and $O(\frac{d_x+d_y}{\epsilon^2}\log{NR})$ for general unnormalized model.

\htitle{DI-COD.} DI-COD applied the Dyadic Interval approach \cite{arasu2004approximate} to Co-Occurring Directions, maintaining a hierarchical structure with $L=\log{\frac{R}{\epsilon}}$ parallel levels, each of which contains a dynamic number of blocks. For $i$-th level, the window is segmented into at most $2^{L-i+1}$ block, and each block maintains a COD sketch. The space cost for DI-COD is $O(\frac{(d_x+d_y)R}{\epsilon}\log^2{\frac{R}{\epsilon}})$. 

\htitle{EH-COD.} Exponential Histogram Co-occurring Directions (EH-COD) combines the Exponential Histograms technique \cite{DatarGIM02} and incorporates the COD algorithm for efficiently approximating matrix multiplication within the sliding window model. The space cost for EH-COD is $O(\frac{d_x+d_y}{\epsilon^2}\log{\epsilon NR})$.

\subsection{Experiments Setup}
\htitle{Datasets.} Experiments are conducted on both synthetic and real-world datasets widely used in matrix multiplication \cite{YaoLCWC24,YeLZ16,MrouehMG17,GhashamiDP14,KangKK20}. All datasets are \emph{unnormalized}.
The details are listed below:
\begin{itemize}[leftmargin=10pt]
\item \textbf{Uniform Random \cite{YaoLCWC24,YeLZ16}.} We generate two random matrices: one of size $2000 \times 10000$ and another of size $1000 \times 10000$. The entries of both matrices are drawn uniformly at random from the interval $[0, 1)$. The window size for this dataset is $N = 4000$.
    
\item \textbf{Random Noisy \cite{MrouehMG17,GhashamiDP14}.} 
We generate the input matrix $\boldsymbol{X}^T = \boldsymbol{SDU} + \boldsymbol{F} / \zeta \in \mathbb{R}^{n \times d_x}$. Here, the term $\boldsymbol{SDU}$ represents an $m$-dimensional signal, while the other part $\boldsymbol{F} / \zeta$ is a Gaussian noise matrix, with scalar parameter $\zeta$ controlling the noise-to-signal ratio.
Specifically, $\boldsymbol{S}\in\mathbb{R}^{n\times m}$ is a random matrix where each entry is drawn from a standard normal distribution. $\boldsymbol{D}\in\mathbb{R}^{m\times m}$ is a diagonal matrix with entries $\boldsymbol{D}_{i,i}=1-(i-1)/m$, and $\boldsymbol{U}\in\mathbb{R}^{m\times d_x}$ is a random rotation which represents the row space of the signal and satisfies that $\boldsymbol{U}^T\boldsymbol{U}=I_m$. $\boldsymbol{F}$ is again a Gaussian matrix with each entries generated i.i.d. from a normal distribution $N(0,1)$. Matrix $\boldsymbol{Y}$ is generated in the same manner as $\boldsymbol{X}$. We set $d_x = 2000$, $d_y=1000$, $m = 400$, $\zeta = 100$, and the window size $N = 4000$.

 \item \textbf{Multimodal Data \cite{MrouehMG17}.} We study the empirical performance of the algorithms in approximating correlation between images and captions. Following \cite{MrouehMG17}, we consider Microsoft COCO dataset \cite{LinMBHPRDZ14}. For visual features we use the residual CNN Resnet101 \cite{HeZRS16} to generate a feature vector of dimension $d_x = 2048$ for each picture. For text we use the Hierarchical Kernel Sentence Embedding \cite{mroueh2015asymmetrically}, resulting in a feature vector of dimensions $d_y = 3000$. We construct the matrices $\boldsymbol{X}$ and $\boldsymbol{Y}$ with sizes $2048 \times 123287$ and $3000 \times 123287$, respectively, where each column represents a feature vector. The window size is set to $N = 10000$.

 \item \textbf{HPMaX \cite{KangKK20}:} We also include the dataset HPMaX, which is used to test the performance of heterogenous parallel algorithms for matrix multiplication. In this dataset, both of $\boldsymbol{X}$ and $\boldsymbol{Y}$  have size of $16384\times 32768$. The window size $N$ is $10000$.
\end{itemize}

\htitle{Evaluation Metrics.} Recall that our \oursolution achieves the optimal space complexity while providing an $\epsilon$-approximation guarantee. Therefore, we design the experiments to explicitly demonstrate the trade-off between space consumption and empirical accuracy across different datasets. Specifically, we tune the parameters of each algorithm and report both the maximum sketch size and the empirical relative correlation error.  
\begin{itemize}[topsep=0.5mm, partopsep=0pt, itemsep=0pt, leftmargin=10pt] 
    \item \textbf{Maximum sketch size}. This metric is measured by the \textit{maximum} number of column vectors maintained by a matrix sketching algorithm. The maximum sketch size metric represents the peak space cost of a matrix sketching algorithm. 
    \item \textbf{Relative correlation error}. This metric is used to assess the approximation quality of the output matrices. It is defined as $\left\| \boldsymbol{X}_W \boldsymbol{Y}_W^T - \boldsymbol{A}_W \boldsymbol{B}_W^T\right\|_2 /\left\|\boldsymbol{X}_W\right\|_F\left\|\boldsymbol{Y}_W\right\|_F$, where $\boldsymbol{X}_W$ and $\boldsymbol{X}_W$ denotes the exact matrices covered by the current window, and $\boldsymbol{A}_W$ and $\boldsymbol{B}_W$ denotes sketch matrices for $\boldsymbol{X}_W\boldsymbol{Y}_W^T$. 
\end{itemize}

\subsection{Experimental Results}
We first adjust the error parameter $\epsilon$ for each algorithm to analyze the trade-off between space efficiency and empirical accuracy. Generally, when the error parameter $\epsilon$ decreases, the maximum sketch size increases. As shown in Figures~\ref{fig:max-error}--\ref{fig:avg-error}, we report the maximum sketch size, as well as the maximum and average relative correlation errors, for each algorithm. Both the x-axis and y-axis are displayed on a logarithmic scale to encompass the wide range of values. 

First, we observe that the curve representing our solution \oursolution consistently resides in the lower-left corner compared to other baselines, in terms of both maximum and average errors.  This implies that for a given space cost (i.e., maximum sketch size), our \oursolution consistently produces matrices with much lower correlation errors. Therefore, our solution demonstrates a superior space-error trade-off, aligning with its optimal space complexity as discussed in Section~\ref{sec:unnormalized-setting}. Second, on certain datasets (e.g., Multimodal Data and HPMax), the second-best algorithm, EH-COD, produces results comparable to our solution when the maximum sketch size is small (i.e., the error parameter $\epsilon$ is large). However, the gap between the two curves widens as the maximum sketch size increases (i.e., the error parameter $\epsilon$ decreases). This is also aligned with the theoretical result that the suboptimal space complexity $O(\frac{d_x + d_y}{\epsilon^2}\log{\epsilon NR})$ of EH-COD is outperformed by our optimal complexity $O(\frac{d_x + d_y}{\epsilon}\log{R})$. Finally, we observe that the EH-COD baseline performs better than the DI-COD baseline in almost all cases, which aligns with the observations in \cite{YaoLCWC24}.

Then, we examine the impact of the error parameter on the space cost of each algorithm. We vary the parameter $\epsilon$ and report the maximum value of sketch size. The results are shown in Figure~\ref{fig:sketch-size}.  The curve of our solution \oursolution consistently remains the lowest. This indicates that, for a given error parameter, \oursolution requires the least space, thereby confirming the conclusion of space optimality. One may note that as $\log_{10}(1/\epsilon)$ increases, the space growth of the EH-COD algorithm gradually slows down. This occurs because, as $\epsilon$ decreases, the storage capacity of EH-COD increases, and the entire sketch becomes sufficient to store the entire window without significant COD compression operations. Consequently, the maximum sketch size approaches the window size.

In summary, when space is the primary concern, our \oursolution is the preferred choice, delivering the best accuracy under space constraints compared to all competitors.

\section{Conclusion}
\label{sec:sec-Conclusion}

In this paper, we propose \oursolution, a novel algorithm for approximate matrix multiplication over sliding windows. \oursolution achieves  the optimal space complexity  in both the normalized and unnormalized settings. Experiments on synthetic and real-world datasets demonstrate its space efficiency and superior estimation accuracy. Future work will focus on reducing the time complexity to achieve linear dependence on the number of non-zero entries.

% \balance
%\begin{acks}
%\end{acks}

%\begin{acks}
%\end{acks}
% \newpage
\bibliographystyle{ACM-Reference-Format}
\bibliography{AMM}

% \newpage

\appendix
\section{Proofs}\label{app:proofs}

\subsection{Proof of Lemma \ref{lem:fast-equivalent}.} 
We note that:

\begin{align}
    \hat{\boldsymbol{A}}{'} & = \hat{\boldsymbol{A}} - \boldsymbol{Q}_X\boldsymbol{u}_i\boldsymbol{u}_i^T\boldsymbol{R}_X \nonumber \\
    & = \boldsymbol{Q}_X(\boldsymbol{R}_X - \boldsymbol{u}_i\boldsymbol{u}_i^T\boldsymbol{R}_X) = \boldsymbol{Q}_X \boldsymbol{R}_X' \nonumber  \\
    \hat{\boldsymbol{B}}{'} & = \hat{\boldsymbol{B}} - \boldsymbol{Q}_Y\boldsymbol{v}_i\boldsymbol{v}_i^T\boldsymbol{R}_Y \nonumber \\
    & = \boldsymbol{Q}_Y(\boldsymbol{R}_Y - \boldsymbol{v}_i\boldsymbol{v}_i^T\boldsymbol{R}_Y) = \boldsymbol{Q}_Y\boldsymbol{R}_Y' \nonumber
\end{align}
Then, we further have that:
\begin{align}
    \hat{\boldsymbol{A}}{'}{\hat{\boldsymbol{B}}{'}}^T & = (\hat{\boldsymbol{A}} - \boldsymbol{Q}_X\boldsymbol{u}_i\boldsymbol{u}_i^T\boldsymbol{R}_X)(\hat{\boldsymbol{B}} - \boldsymbol{Q}_Y\boldsymbol{v}_i\boldsymbol{v}_i^T\boldsymbol{R}_Y)^T \nonumber \\ 
    & = \hat{\boldsymbol{A}}\hat{\boldsymbol{B}}^T - \boldsymbol{Q}_X\boldsymbol{u}_i\boldsymbol{u}_i^T\boldsymbol{R}_X\hat{\boldsymbol{B}}^T - \hat{\boldsymbol{A}}\boldsymbol{R}_Y^T\boldsymbol{v}_i\boldsymbol{v}_i^T\boldsymbol{Q}_Y^T \nonumber \\
    &\ \ \ \ + \boldsymbol{Q}_X\boldsymbol{u}_i\boldsymbol{u}_i^T\boldsymbol{R}_X\boldsymbol{R}_Y^T\boldsymbol{v}_i\boldsymbol{v}_i^T\boldsymbol{Q}_Y^T \nonumber \\
    & = \hat{\boldsymbol{A}}\hat{\boldsymbol{B}}^T - \boldsymbol{Q}_X\boldsymbol{u}_i\boldsymbol{u}_i^T\boldsymbol{R}_X\boldsymbol{R}_Y^T\boldsymbol{Q}_Y^T - \boldsymbol{Q}_X \boldsymbol{R}_X
    \boldsymbol{R}_Y^T\boldsymbol{v}_i\boldsymbol{v}_i^T\boldsymbol{Q}_Y^T \nonumber \\
    &\ \ \ \ + \boldsymbol{Q}_X\boldsymbol{u}_i\boldsymbol{u}_i^T\boldsymbol{R}_X\boldsymbol{R}_Y^T\boldsymbol{v}_i\boldsymbol{v}_i^T\boldsymbol{Q}_Y^T \nonumber \\
    & = \hat{\boldsymbol{A}}\hat{\boldsymbol{B}}^T - \boldsymbol{Q}_X\boldsymbol{u}_i\boldsymbol{u}_i^TU\Sigma V^T\boldsymbol{Q}_Y^T - \boldsymbol{Q}_XU\Sigma V^T\boldsymbol{v}_i\boldsymbol{v}_i^T\boldsymbol{Q}_Y^T \nonumber \\
    &\ \ \ \ + \boldsymbol{Q}_X\boldsymbol{u}_i\boldsymbol{u}_i^TU\Sigma V^T\boldsymbol{v}_i\boldsymbol{v}_i^T\boldsymbol{Q}_Y^T \nonumber \\
    & = \hat{\boldsymbol{A}}\hat{\boldsymbol{B}}^T - \boldsymbol{Q}_X\boldsymbol{u}_i\sigma_i\boldsymbol{v}_i^T\boldsymbol{Q}_Y^T \nonumber \\
    & = \hat{\boldsymbol{A}}\hat{\boldsymbol{B}}^T - \hat{a_i}\hat{b_i}^T \nonumber
\end{align}
This finishes the proof. 

\subsection{Proof of Theorem \ref{thm:socod-normalized}.}

We denote the matrices of snapshot vectors generated between timestamps $a$ and $b$ as $\boldsymbol{C}_{a,b}$ and $\boldsymbol{D}_{a,b}$. Specifically, $\boldsymbol{C}_{a,b}$ if formed by stacking $s_j.u$ for all $s_j.t\in [a,b]$ and $\boldsymbol{D}_{a,b}$ is formed by stacking $s_j.v$ for all $s_j.t\in [a,b]$. Similarly, let $\boldsymbol{X}_{a,b}$ and $\boldsymbol{Y}_{a,b}$ represent the matrices stacked by all $\boldsymbol{x}_i$ and $\boldsymbol{y}_i$, respectively, where $i\in[a,b]$. We denote $P$ as the moment right before the primary sketch begins to receive updates. We have $P = (k-1)N$, where $k = \max(1,\lfloor (T-1)/N\rfloor)$. And we define $(\hat{\boldsymbol{A}}_T,\hat{\boldsymbol{B}}_T)$ as the primary COD sketch at moment $T$. Next, we have

\begin{align}
    & \left\| \boldsymbol{X}_W\boldsymbol{Y}_W^T - \boldsymbol{A}_{aug}\boldsymbol{B}_{aug}^T\right\|_2 \nonumber \\
    & = \left\| \boldsymbol{X}_{T-N+1,T}\boldsymbol{Y}_{T-N+1,T}^T - \hat{\boldsymbol{A}}_T\hat{\boldsymbol{B}}_T^T - \boldsymbol{C}_{T-N+1,T}\boldsymbol{D}_{T-N+1,T}^T\right\|_2 \nonumber \\
    & = || \boldsymbol{X}_{P,T}\boldsymbol{Y}_{P,T}^T - \boldsymbol{X}_{P,T-N}\boldsymbol{Y}_{P,T-N}^T - \hat{\boldsymbol{A}}_T\hat{\boldsymbol{B}}_T^T \nonumber \\
    & - \boldsymbol{C}_{P,T}\boldsymbol{D}_{P,T}^T + \boldsymbol{C}_{P,T-N}\boldsymbol{D}_{P,T-N}^T ||_2 \nonumber \\
    & \leq \left\| \boldsymbol{X}_{P,T}\boldsymbol{Y}_{P,T}^T - \boldsymbol{C}_{P,T}\boldsymbol{D}_{P,T}^T - \hat{\boldsymbol{A}}_T\hat{\boldsymbol{B}}_T^T \right\| + \nonumber \\
    & \left\| \boldsymbol{X}_{P,T-N}\boldsymbol{Y}_{P,T-N}^T - \boldsymbol{C}_{P,T-N}\boldsymbol{D}_{P,T-N}^T \right\|_2 \nonumber
\end{align}

\begin{align}
    & = \left\| [\boldsymbol{X}_{P,T},-\boldsymbol{C}_{P,T}][\boldsymbol{Y}_{P,T},\boldsymbol{D}_{P,T}]^T - \hat{\boldsymbol{A}}_T\hat{\boldsymbol{B}}_T^T \right\| \nonumber \\
    & + \left\| \boldsymbol{X}_{P,T-N}\boldsymbol{Y}_{P,T-N}^T - \boldsymbol{C}_{P,T-N}\boldsymbol{D}_{P,T-N}^T - \hat{\boldsymbol{A}}_{T-N}{\hat{\boldsymbol{B}}_{T-N}}^T + \hat{\boldsymbol{A}}_{T-N}{\hat{\boldsymbol{B}}_{T-N}}^T \right\|_2 \nonumber \\
    & \leq \left\| [\boldsymbol{X}_{P,T},-\boldsymbol{C}_{P,T}][\boldsymbol{Y}_{P,T},\boldsymbol{D}_{P,T}]^T - \hat{\boldsymbol{A}}_T\hat{\boldsymbol{B}}_T^T \right\| + \left\|\hat{\boldsymbol{A}}_{T-N}{\hat{\boldsymbol{B}}_{T-N}}^T \right\|_2 \nonumber \\
    & + \left\| [\boldsymbol{X}_{P,T-N},- \boldsymbol{C}_{P,T-N}][\boldsymbol{Y}_{P,T-N},\boldsymbol{D}_{P,T-N}]^T  - \hat{\boldsymbol{A}}_{T-N}{\hat{\boldsymbol{B}}_{T-N}}^T\right\|  \nonumber \\
\end{align}

where $(\hat{\boldsymbol{A}}_{T-N},\hat{\boldsymbol{B}}_{T-N})$ and $(\hat{\boldsymbol{A}}_T,\hat{\boldsymbol{B}}_T)$ can be considered as the COD sketches of matrices $([\boldsymbol{X}_{P,T-N},- \boldsymbol{C}_{P,T-N}], [\boldsymbol{Y}_{P,T-N},\boldsymbol{D}_{P,T-N}])$ and $([\boldsymbol{X}_{P,T},-\boldsymbol{C}_{P,T}],$ $[\boldsymbol{Y}_{P,T},\boldsymbol{D}_{P,T}])$, respectively. Extracting a snapshot with vectors $(v,u)$ from $(\hat{\boldsymbol{A}},\hat{\boldsymbol{B}})$ can be seen as a COD sketch update with pair of columns $(-v,u)$. It is also known that $\left\|\boldsymbol{C}_{P,T} \right\|_F\left\|\boldsymbol{D}_{P,T} \right\|_F\leq \left\|\boldsymbol{X}_{P,T}\boldsymbol{Y}_{P,T}^T\right\|_*\leq \left\|\boldsymbol{X}_{P,T} \right\|_F\left\|\boldsymbol{Y}_{P,T} \right\|_F = T - P$. Similarly, we also have $\left\|\boldsymbol{C}_{P,T-N} \right\|_F\left\|\boldsymbol{D}_{P,T-N} \right\|_F\leq T - N - P$.

By the approximation error guarantee of COD \cite{MrouehMG17}, we have:
    \begin{align}
        & \left\| [\boldsymbol{X}_{P,T},-\boldsymbol{C}_{P,T}][\boldsymbol{Y}_{P,T},\boldsymbol{D}_{P,T}]^T - \hat{\boldsymbol{A}}_T\hat{\boldsymbol{B}}_T^T \right\|_2 \nonumber \\
        & \leq \epsilon \left\| [\boldsymbol{X}_{P,T},-\boldsymbol{C}_{P,T}]\right\|_F\left\| [\boldsymbol{Y}_{P,T},-\boldsymbol{D}_{P,T}]\right\|_F = 2\epsilon(T-P) \nonumber \\
        & \left\| [\boldsymbol{X}_{P,T-N},-\boldsymbol{C}_{P,T-N}][\boldsymbol{Y}_{P,T-N},\boldsymbol{D}_{P,T-N}]^T - \hat{\boldsymbol{A}}_{T-N}\hat{\boldsymbol{B}}_{T-N}^T \right\|_2 \nonumber \\
        & \leq \epsilon \left\| [\boldsymbol{X}_{P,T-N},-\boldsymbol{C}_{P,T-N}]\right\|_F\left\| [\boldsymbol{Y}_{P,T-N},-\boldsymbol{D}_{P,T-N}]\right\|_F = 2\epsilon(T-N-P) \nonumber \\
        & \left\|\hat{\boldsymbol{A}}_{T-N}{\hat{\boldsymbol{B}}_{T-N}}^T \right\|_2\leq \epsilon\left\|\boldsymbol{X}_{P,T-N} \right\|_F \left\|\boldsymbol{Y}_{P,T-N} \right\|_F = \epsilon(T - N - P)
    \end{align}
\noindent
Combining Inequality (2), the approximation error is bounded by

\begin{align}
    & \left\| \boldsymbol{X}_W\boldsymbol{Y}_W^T - \boldsymbol{A}_{aug}\boldsymbol{B}_{aug}^T\right\|_2 \nonumber \\
    & \leq \left\| [\boldsymbol{X}_{P,T},-\boldsymbol{C}_{P,T}][\boldsymbol{Y}_{P,T},\boldsymbol{D}_{P,T}]^T - \hat{\boldsymbol{A}}_T\hat{\boldsymbol{B}}_T^T \right\| + \left\|\hat{\boldsymbol{A}}_{T-N}{\hat{\boldsymbol{B}}_{T-N}}^T \right\|_2 \nonumber \\
    & + \left\| [\boldsymbol{X}_{P,T-N},- \boldsymbol{C}_{P,T-N}][\boldsymbol{Y}_{P,T-N},\boldsymbol{D}_{P,T-N}]^T  - \hat{\boldsymbol{A}}_{T-N}{\hat{\boldsymbol{B}}_{T-N}}^T\right\|  \nonumber \\
    & \leq 2\epsilon(T-P) + 2\epsilon(T-N-P) + \epsilon(T-N-P) \nonumber \\
    & \leq \epsilon(5T - 5P - 3N)  \leq 8\epsilon N
\end{align}

    Next, we prove the number of snapshots is bounded by $O(\frac{1}{\epsilon})$. Suppose that at time $T$ the queue has $w$ snapshots of vector pairs $(\boldsymbol{u}_1,\boldsymbol{v}_1),(\boldsymbol{u}_2,\boldsymbol{v}_2),\cdots,(\boldsymbol{u}_k,\boldsymbol{v}_k)$, and the corresponding singular values are $\sigma_1,\sigma_2,\cdots,\sigma_k$. Then we know $k\epsilon N\leq $$\sum_{i=1}^k\left\|\boldsymbol{u}_i\right\|_2\left\|\boldsymbol{v}_i\right\|_2$$=\sum_{i=1}^k\sigma_i\leq \left\|\hat{\boldsymbol{A}}_{T-N}{\hat{\boldsymbol{B}}_{T-N}}^T\right\|_{*}$$ + \left\|\boldsymbol{X}_W\boldsymbol{Y}_W^T\right\|_* \leq l\epsilon N + N = 2N$. Hence, we have $k\epsilon N \leq 2N$, which implies $k\leq \frac{2}{\epsilon}$. Thus, the space cost of \oursolution for normalized model is dominated by the COD sketch and snapshot storage, yielding a total space cost of $O((d_x+d_y)/\epsilon)$.

 \subsection{Proof of Theorem \ref{thm:socod-unnormalized}.}

    For $i$-th level of \newsolution, the correlation error is $2^{i+3}\epsilon N$ by the theorem \ref{thm:socod-normalized}. To ensure the correlation error $\left\| \boldsymbol{X}_W\boldsymbol{Y}_W^T - \boldsymbol{A}_{aug}\boldsymbol{B}_{aug}^T\right\|_2$ is bounded by $4\epsilon \left\| \boldsymbol{X}_W\right\|_F\left\| \boldsymbol{Y}_W\right\|_F$, we need to find the level $i$ such that $2^{i+3}\epsilon N \leq 4\epsilon \left\| \boldsymbol{X}_W\right\|_F\left\| \boldsymbol{Y}_W\right\|_F < 2^{i+4}\epsilon N $, thus we have

    \begin{align}
        log_2{\frac{\left\| \boldsymbol{X}_W\right\|_F\left\| \boldsymbol{Y}_W\right\|_F}{2N}} - 1 < i \leq  log_2{\frac{\left\| \boldsymbol{X}_W\right\|_F\left\| \boldsymbol{Y}_W\right\|_F}{2N}} \label{ine:level-i-range}
    \end{align}

    At the same time, we must ensure level $i$ contains sufficient snapshots. Suppose at time $T$, the snapshot queue $M[i].S$ holds $k$ snapshots $(u_1,v_1),(u_2,v_2),\cdots,(u_k,v_k)$. Given the register threshold $\theta = 2^i\epsilon N$, we have $\sum_{i=1}^k\left\|u_i \right\|_2\left\|v_i\right\|_2 \geq k\times2^i\epsilon N$. This sum consists of contributions from both the residue sketch matrix and the term $\sum_{i=T-N+1}^T \left\| \boldsymbol{x}_i \boldsymbol{y}_i^T\right\|_*$. From this, we derive the upper bound:
    
    \begin{align}
      &  k\times2^i\epsilon N  \leq \sum_{i=1}^k\left\|u_i \right\|_2\left\|v_i\right\|_2  \leq \sum_{i=T-N+1}^T \left\| \boldsymbol{x}_i \boldsymbol{y}_i^T\right\|_* + l2^i\epsilon N  \nonumber \\
        & = 2^iN + \sum_{i=T-N+1}^T \left\| \boldsymbol{x}_i\right\|_2 \left\| \boldsymbol{y}_i\right\|_2  \leq 2^iN + \left\| \boldsymbol{X}_W\right\|_F \left\| \boldsymbol{Y}_W\right\|_F \label{ine:snapshot-comtribute}
    \end{align}
    
    Combining (\ref{ine:level-i-range}) and (\ref{ine:snapshot-comtribute}), we deduce that $k\times2^i\epsilon N < 6\times 2^iN$, which implies $k<\frac{6}{\epsilon}$. As a result, storing at most $\frac{6}{\epsilon}$ snapshots per level suffices. Thus, \newsolution derives a space complexity of $O(\frac{d_x+d_y}{\epsilon}\log{R})$.

\subsection{Proof of Theorem \ref{thm:lower-bound-amm-sw}.} Without loss of generality, we suppose that $d_y\geq d_x$. We partition a window of size $N$ consisting of $(d_x+1)$-
dimensional $\boldsymbol{X}_W$ and $(d_y+1)$-dimensional $\boldsymbol{Y}_W$ into $\log{R}+ 2$ blocks, and we label the leftmost $\log{R}+1$ blocks as $\log{R},\cdots,2,1,0$ from left to right. The construction of these $\log{R}+2$ blocks are as follow: 

\begin{enumerate}[leftmargin=16pt]
    \item Choose $\log{R}+1$ matrix pairs from the set $\hat{\mathcal{Z}}_{l/4}$ from the lemma \ref{lem:cardinality}, and by setting $\delta = 1/8$, we have $M = 2^{\Omega(d_yl)}$ and $\left\|\boldsymbol{X}_i\boldsymbol{Y}_i^T-\boldsymbol{X}_j\boldsymbol{Y}_j^T\right\|_2 > 1/2$ for any $i\neq j$ and $i,j\in[1,M]$. Thereby the total number of distinct arrangements $L = \binom{2^{\Omega(d_yl)}}{\log{R}+1}$, consequently resulting that $\log{L} = \Omega(d_yl\log{R})$.

    \item For block $i$, multiply the chosen $(\boldsymbol{X}_i\in \mathbb{R}^{d_x\times (l/4)},\boldsymbol{Y}_i\in \mathbb{R}^{d_y\times (l/4)})$ by a scalar of $\sqrt{\frac{2^iN}{l}}$, making $\left\|\boldsymbol{X}_i\right\|_F\left\|\boldsymbol{Y}_i\right\|_F = 2^{i-2}N$.

    \item For block $i$ where $i>\log{\frac{lR}{N}}$, we replace the scalar $\sqrt{\frac{2^iN}{l}}$ with $\sqrt{R}$, and increases the number of columns from $l/4$ to $(l/4)\cdot 2^{i - \log{\frac{lR}{N}}}$, ensuring $1\leq\left\|\boldsymbol{x}_i\right\|_2^2, \left\|\boldsymbol{y}_i\right\|_2^2\leq R$. This configuration preserves the condition $\left\|\boldsymbol{X}_i\right\|_F\left\|\boldsymbol{Y}_i\right\|_F = 2^{i-2}N$. To bound the total number of columns by $N$, the condition $H = \frac{l}{4}\log{\frac{lR}{N}} +  \sum_{i = \log{\frac{lR}{N}}}^{\log{R}} \frac{l}{4}\cdot 2^{i - \log{\frac{lR}{N}}} \leq N$ must hold, implying $N \geq \frac{l}{2}\log{\frac{lR}{N}}$. 

    \item If the total number of columns $H$ of these blocks is less than $N$, to fill up the whole window, we set the last ($\log{R}+2$)-th block as all zeros matrix pair $(\boldsymbol{X}_{\log{R}+2},\boldsymbol{Y}_{\log{R}+2})=(0^{d_x\times(N-H)},0^{d_y\times(N-H)})$.

    \item Add one dimension to all matrices in all $\log{R}+2$ blocks and set the newly-added ($d_x+1$)-th dimension of each $\boldsymbol{X}_i$ and ($d_y+1$)-th dimension of each $\boldsymbol{Y}_i$ as $1$.
    
\end{enumerate}

We assume the algorithm receives one of these $L$ predefined arrangements of length $N$, followed by a sequence of $N$ one-hot vector pairs. In these pairs, only the ($d_x+1$)-th dimension in $\boldsymbol{X}$ and the ($d_y+1$)-dimension in $\boldsymbol{Y}$ are set to $1$. Let $(\boldsymbol{X}_W^i, \boldsymbol{Y}_W^i)$ denote the matrices over the sliding window of length $N$ when $\log{R},\cdots,i+1,i$ blocks have exactly expired. Assume the algorithm can provide AMM estimation $(\boldsymbol{A}_W^i, \boldsymbol{B}_W^i)$ and $(\boldsymbol{A}_W^{i-1}, \boldsymbol{B}_W^{i-1})$ for $(\boldsymbol{X}_W^i, \boldsymbol{Y}_W^i)$ and $(\boldsymbol{X}_W^{i-1}, \boldsymbol{Y}_W^{i-1})$, respectively, with a correlation coefficient $\frac{1}{6l}$, implying:
$$
\left\|\boldsymbol{X}_W^i {\boldsymbol{Y}_W^i}^T - \boldsymbol{A}_W^i {\boldsymbol{B}_W^i}^T\right\|_2
\leq \frac{1}{6l}
\left\|\boldsymbol{X}_W^i\right\|_F\left\|\boldsymbol{Y}_W^i\right\|_F = \frac{1}{6l}(\frac{N}{4}\cdot 2^{i+1}+\frac{3N}{4}),
$$

$$
\left\|\boldsymbol{X}_W^{i-1} {\boldsymbol{Y}_W^{i-1}}^T - \boldsymbol{A}_W^{i-1} {\boldsymbol{B}_W^{i-1}}^T\right\|_2
\leq \frac{1}{6l}
\left\|\boldsymbol{X}_W^{i-1}\right\|_F\left\|\boldsymbol{Y}_W^{i-1}\right\|_F = \frac{1}{6l}(\frac{N}{4}\cdot 2^{i}+\frac{3N}{4})
$$

Then We can compute the AMM for block $i$ using $(\boldsymbol{A}_i',\boldsymbol{B}_i')$, where $\boldsymbol{A}_i'{\boldsymbol{B}_i'}^T = \boldsymbol{A}_W^{i} {\boldsymbol{B}_W^{i}}^T - \boldsymbol{A}_W^{i-1} {\boldsymbol{B}_W^{i-1}}^T$ as described below:

\begin{align}
    & \left\|\boldsymbol{X}_i\boldsymbol{Y}_i^T- \boldsymbol{A}_i'{\boldsymbol{B}_i'}^T\right\|_2 \nonumber \\
    = & \left\|(\boldsymbol{X}_W^i {\boldsymbol{Y}_W^i}^T - \boldsymbol{A}_W^i {\boldsymbol{B}_W^i}^T) - (\boldsymbol{X}_W^{i-1} {\boldsymbol{Y}_W^{i-1}}^T - \boldsymbol{A}_W^{i-1} {\boldsymbol{B}_W^{i-1}}^T)\right\|_2 \nonumber \\
    \leq & \left\|\boldsymbol{X}_W^i {\boldsymbol{Y}_W^i}^T - \boldsymbol{A}_W^i {\boldsymbol{B}_W^i}^T\right\|_2 + \left\|\boldsymbol{X}_W^{i-1} {\boldsymbol{Y}_W^{i-1}}^T - \boldsymbol{A}_W^{i-1} {\boldsymbol{B}_W^{i-1}}^T\right\|_2 \nonumber \\
    \leq & \frac{1}{6l}(\frac{3}{4}\cdot 2^i+ \frac{3}{2}N) \leq \frac{1}{l}\left\|\boldsymbol{X}_i\right\|_F\left\|\boldsymbol{Y}_i\right\|_F \nonumber
\end{align}

The algorithm can estimate blocks for all layers $0\leq i\leq \log{R}$. By lemma \ref{lem:lower-bound-amm}, estimating of each block requires $\Omega((d_x+d_y)l)$ bits of space. Consequently, we establish the fundamental space complexity lower bound for any deterministic AMM algorithm over a sliding window as $\Omega(\frac{d_x+d_y}{\epsilon}\log{R})$. 

In the special case $X=Y$, the problem reduces to matrix sketching over a sliding window, for which the lower bound $\Omega(\frac{d}{\epsilon}\log{R})$ has been established in \cite{YinWLWZHL24}, consistent with our findings.

\section{Additional Examples}\label{app:examples}

Here, we provide an example of how the $\lambda$-snapshot method works for counting the number of 1-bits in a sliding window.
\begin{example}
Consider a bit stream 
\(
\{1,0,1,1,0,1,1,1,0,1,0\}
\)
with 11 elements and let \(\lambda=3\). The 1-bits occur at positions 1, 3, 4, 6, 7, 8, and 10; hence, the 3\textsuperscript{rd} and 6\textsuperscript{th} 1-bits (positions 4 and 8) are sampled. We partition the stream into \(\lambda\)-blocks of length 3:
\[
\text{Block 1: } [1,3],\quad \text{Block 2: } [4,6],\quad \text{Block 3: } [7,9],\quad \text{Block 4: } [10,12].
\]
(Note that Block 4 is still incomplete.) At time \(T=11\), assume a sliding window of size \(N=5\):
\(
W_{11} = [7,11].
\)
Within \(W_{11}\), the 1-bits occur at positions 7, 8, and 10. Block 2 lies entirely outside \(W_{11}\) and is dropped, and although Block 4 overlaps \(W_{11}\), it contains no sampled bit. Thus, only Block 3 (spanning 7–9 and containing the sampled 1-bit at position 8) remains registered in the queue \(Q\). Moreover, the 1-bit at position 10, arriving after the last sampled bit, yields a leftover count \(\ell=1\). The \(\lambda\)-snapshot estimate is hence
\(
v(S)=|Q|\cdot\lambda + \ell = 1\cdot3 + 1 = 4.
\)
Since the true count of 1-bits in \(W_{11}\) is 3, the error is \(4-3=1\).
\end{example}

\end{document}